\documentclass[twocolumn]{aastex631}
\usepackage{makecell}

\begin{document}

\title{Physical Pathways for JWST-Observed Supermassive Black Holes in the Early Universe}

\author[0000-0002-6038-5016]{Junehyoung Jeon}
\affiliation{Department of Astronomy, University of Texas, Austin, TX 78712, USA}
\author[0000-0003-0212-2979]{Volker Bromm}
\affiliation{Department of Astronomy, University of Texas, Austin, TX 78712, USA}
\affiliation{Weinberg Institute for Theoretical Physics, University of Texas, Austin, TX 78712, USA}
\author[0000-0002-4966-7450]{Boyuan Liu}
\affiliation{Institute of Astronomy, University of Cambridge, Cambridge CB3 0HA, UK}
\affiliation{Institut für Theoretische Astrophysik, Zentrum für Astronomie, Universität Heidelberg, D-69120 Heidelberg, Germany}
\author[0000-0001-8519-1130]{Steven L.~Finkelstein}
\affiliation{Department of Astronomy, University of Texas, Austin, TX 78712, USA}
\email{junehyoungjeon@utexas.edu}

\begin{abstract}
Observations with the \textit{James Webb Space Telescope} (\textit{JWST}) have revealed active galactic nuclei (AGN) powered by supermassive black holes (SMBHs) with estimated masses of $10^7-10^8$ M$_\odot$ at redshifts $z\sim7-9$. Some reside in overmassive systems with higher AGN to stellar mass ratios than locally. Understanding how massive black holes could form so early in cosmic history and affect their environment to establish the observed relations today are some of the major open questions in astrophysics and cosmology. One model to create these massive objects is through direct collapse black holes (DCBHs) that provide massive seeds ($\sim10^5-10^6$ M$_\odot$), able to reach high masses in the limited time available. We use the cosmological simulation code GIZMO to study the formation and growth of DCBH seeds in the early Universe. To grow the DCBHs, we implement a gas swallowing model set to match the Eddington accretion rate as long as the nearby gaseous environment, affected by stellar and accretion disk feedback, provides sufficient fuel. We find that to create massive AGN in overmassive systems at high redshifts, massive seeds accreting more efficiently than the fiducial Bondi-Hoyle model are needed. We assess whether the conditions for such enhanced accretion rates are realistic by considering limits on plausible transport mechanisms. We also examine various DCBH growth histories and find that mass growth is more sustained in overdense cosmological environments, where high gas densities are achieved locally. We discuss the exciting prospect to directly probe the assembly history of the first SMBHs with upcoming, ultra-deep \textit{JWST} surveys.  
\end{abstract}

\keywords{Early universe — Galaxy formation — Supermassive black holes — 
Active galactic nuclei — Theoretical models}

\section{Introduction} \label{sec:intro}
How did accreting supermassive black holes (SMBHs) or active galactic nuclei (AGN) impact early cosmic history \citep{Woods2019}? The expectation is that they were abundant in the high-$z$ Universe, to be detected in ongoing and future surveys \citep{Volonteri2017,Li2023,Spurzem2023,Evans2023}, unless they are obscured or too faint to be discovered \citep{Smith2019,Gilli2022,Goulding2022,Jeon2023}. The recent influx of new observations with the \textit{James Webb Space Telescope} (\textit{JWST}) confirmed such predictions and revealed a large number of active galactic nuclei (AGN) in the early Universe \citep[e.g.][]{Kocevski2023,Ding2022,Larson2023,Onoue2023,Furtak2023,Bogdan2023,Juodbalis2023,Bosman2023,Greene2023,Kokorev2023,Fujimoto2023,Maiolino2023}. 

These new observations help reveal the nature of early cosmic evolution, as AGN influence their environment and host galaxies through the large amounts of energy released. In the local Universe, we observe relationships between the AGN and host galaxy/halo properties, including the SMBH mass and stellar velocity dispersion ($M-\sigma$) \citep{Gebhardt2000,Graham2011,Kormendy2013}, SMBH mass and galaxy luminosity \citep{Beifiori2012}, or between the masses of the SMBH and stellar component \citep{Croton2006,Ding2020}. Such observed relations further suggest that AGN and their host galaxies co-evolve \citep{Heckman2014}. When and how these relationships were established can be investigated with the new high-redshift observations \citep[e.g.][]{Pacucci2023,Wang2023,Yue2023,Thomas2023}. Moreover, cosmological simulations and semi-analytic models have shown that AGN feedback could quench star formation (SF) through heating the interstellar medium (ISM) and ejecting the gas \citep{Dubois2013,Wagner2016,Shirakata2019}, or enhance SF by pressurizing the ISM gas \citep{Valentini2021,Mercedes2023} which may lead to the observed local relations.

UV emission from AGN could have also been a significant contributor to reionization. The general view has been that the AGN number density is too low at high redshifts to complete reionization compared to UV radiation from galactic sources \citep[e.g.][]{Ciardi2003,Robertson2015,Parsa2018,Finkelstein2019,Naidu2020}, but some studies have suggested that AGN and stellar sources were comparable in their contribution to reionization \citep[e.g.,][]{Madau2015,Volonteri2016,Jeon2022}. The ongoing discovery of numerous high-redshift AGN may imply that they could indeed have had non-negligible effects on reionization \citep{Fujimoto2023}, if AGN-produced ionizing radiation can escape to the intergalactic medium (IGM). Future observations of additional high redshift sources will allow us to better constrain the overall AGN contribution to reionization.

Furthermore, recent pulsar timing array (PTA) observations have revealed the stochastic gravitational wave background (GWB) pervading the Universe \citep{Agazie2023,Antoniadis2023,Reardon2023,Xu2023}. One of the main proposed sources of the inferred low-frequency GWB are binary SMBHs \citep{Hobbs2017,Romano2017}, and the PTA observations have resulted in new constraints on the binary SMBH population. The constraints are currently weak due to the initial GWB data having low signal-to-noise, but will be improved upon with future observations \citep[e.g.,][]{Agazie2023_2,Agazie2023_3}.

However, despite the numerous ways AGN/SMBHs could contribute to early cosmic evolution, key questions remain regarding how the first SMBHs formed and evolved to what we observe today. It is unclear how SMBHs grew to become the massive $\gtrsim10^9$ M$_\odot$ quasars present already so early in cosmic history at $z\gtrsim 6$ \citep[e.g.,][]{Wu2015,Banados2018,Zubovas2021,Fan2022}. The new \textit{JWST} observations provided more examples of such massive quasars in the young Universe ($z\sim7-10$), extending to slightly less massive SMBHs ($\sim10^7-10^8$ M$_\odot$) that could be the progenitors of the previously known quasars \citep[e.g.,][]{Larson2023,Furtak2023,Greene2023}. The fiducial formation channel, involving a stellar remnant seed and subsequent growth at the theoretical maximum (Eddington-limited) rate, will not have had enough time to form these massive SMBHs \citep{Smith2019_2,Inayoshi2020,Larson2023}, posing a crucial challenge to our understanding of early structure formation. 

To alleviate this bottleneck, different seeding mechanisms have been proposed for SMBH seeds\footnote{Alternatively, the Universe may have been endowed with primordial black holes extending into the SMBH range \citep[e.g.,][]{Carr2020,LiuBromm2022}.}, with two main channels \citep[e.g.,][]{Haemmerle2020,Volonteri2021,Sassano2021}. The first channel originates from the remnants of the first generation of metal-free, Population~III (Pop~III), stars \citep{Madau2001,Heger2003}. They are predicted to have a top-heavy initial mass function (IMF), favoring massive stars \citep[e.g.,][]{Stacy2016,Hirano2017,Latif2022}. The Pop III stellar remnant black holes thus may be more massive than local stellar mass black holes, with $\sim10^2-10^3$ M$_\odot$. After formation, the Pop III seed black holes will grow further through accretion and mergers \citep{Jeon2012,SmithRegan2018,Bhowmick2023}. 

The second, less common channel produces more massive seeds, of order $\sim10^5$ M$_\odot$. These so-called direct collapse black holes (DCBHs) originate from the collapse of massive primordial gas clouds, under rare conditions that allow the gas to collapse without fragmenting to form multiple stars. Instead, the gas cloud will directly become a massive black hole or form a supermassive star that soon collapses into a massive black hole \citep[e.g.,][]{Bromm2003,Begelman2006,Lodato2006,Lodato2007,Johnson2013,Wise2019,Haemmerl2018,Haemmerle2020}. The massive DCBH seeds may also form through runaway collisions of (proto-)stars or mergers of intermediate mass black holes (IMBHs) in dense stellar clusters with high gas inflow rates and rapid accretion flows \citep{Zwick2023,Reinoso2023,Klessen2023,Mayer2024}. Starting with a larger initial mass, the DCBH seeds are expected to have weaker time constraints in growing to the observed high redshift SMBH masses.

\begin{figure*}[htb!]
\gridline{
 \fig{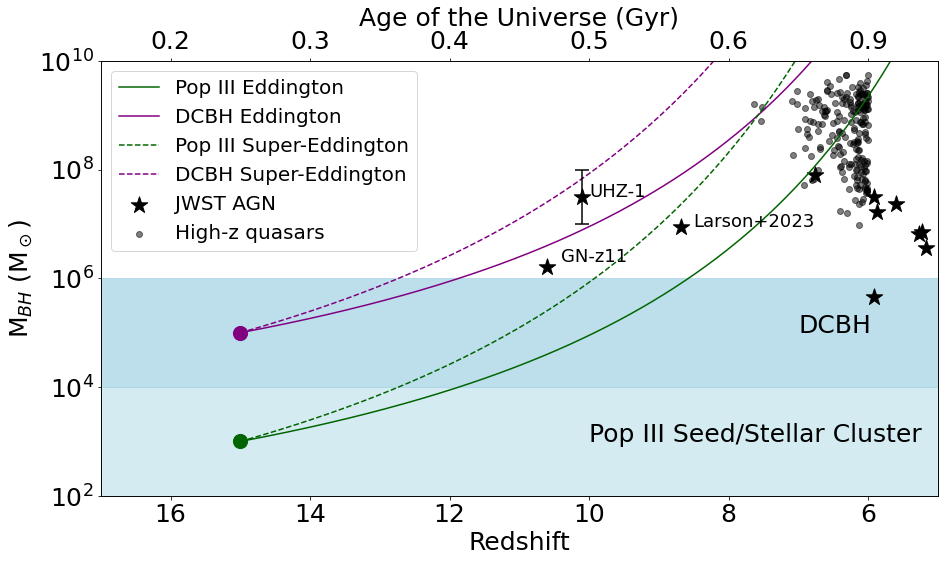}{0.5\textwidth}{}
 \fig{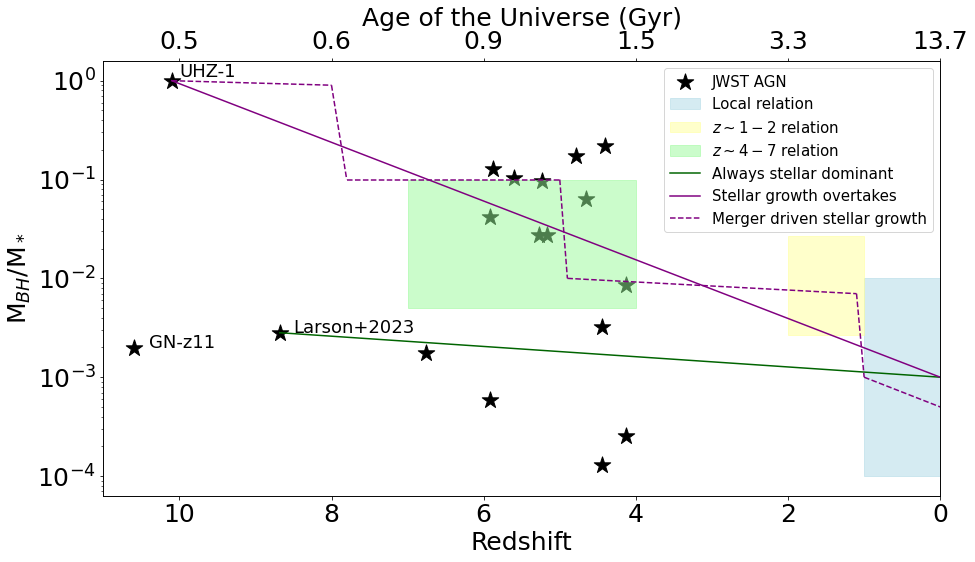}{0.5\textwidth}{}
}    \caption{SMBH growth through cosmic time. {\it Left:} Black hole mass versus redshift or age of the Universe. We show select \textit{JWST} observations of high-redshift AGN \citep{Larson2023,Bogdan2023,Maiolino2023_2}, as well as high-redshift quasars \citep{Wang2010,Willott2017,Decarli2018,Izumi2018,Pensabene2020,Inayoshi2020}. We also indicate in shaded regions the initial black hole seed masses predicted for the Pop~III stellar seed/stellar cluster and DCBH models. To form the highest-redshift AGN and the most massive quasars, the stellar seed model requires growth at super-Eddington rates (1.5 times the Eddington rate for the lines shown in the plot), and the DCBH model at the Eddington rate if they form at $z\sim15$ with a typical seed mass. The necessary conditions could be achieved in the most extreme and dense regions of the Universe. {\it Right:} Black hole mass to host stellar mass ratio versus redshift or age of the Universe. We again show select \textit{JWST} high-redshift AGN observations in comparison to empirical $M_{\rm BH}-M_{*}$ relations inferred for $z\sim0$ \citep{Reines2015}, $z\sim1-2$ \citep{Ding2020}, and $z\sim4-7$ \citep{Pacucci2023}. An initially overmassive system, as suggested for some {\it JWST} sources, could slowly evolve onto the local relation, experiencing a more rapid stellar than black hole growth. Alternatively, the overmassive systems could evolve into the local ones through a series of mergers that significantly increase the stellar mass in short episodes.}
    \label{fig:intro}
\end{figure*}
 
Moreover, the new \textit{JWST} AGN observations, while uncertainties exist in the mass measurements, include cases of ``overmassive" black holes where the SMBH to galaxy stellar mass ratio is much higher than in the local Universe \citep[e.g.,][]{Bogdan2023,Kokorev2023,Pacucci2023,Natarajan2023}. This is consistent with the DCBH formation scenario, where the seeds will form in initially star-poor environments. The stellar/IMBH collision scenario to form heavy seeds can also produce such overmassive systems as the dense stellar clusters where the collisions occur compose a high fraction of the stellar mass of the high redshift galaxies \citep{Mayer2024}. Soon after, the AGN feedback triggers SF in the nearby environment, forming a galaxy with an overmassive SMBH \citep{Wise2019,Agarwal2013}. We schematically summarize the challenge of early SMBH growth in Fig.~\ref{fig:intro}. Extreme conditions such as DCBH and Eddington/super-Eddington growth are necessary to produce high-redshift AGN and quasars, and the overmassive systems will have to evolve onto the local empirical relations through rapid stellar growth and/or mergers with other systems. 

We have explored the Pop III seeding channel in previous work using cosmological simulations \citep{Jeon2023}. However, under the standard seeding and growth models, the black hole seeds did not grow to the observed large masses but remained as low-mass, faint objects below current detection thresholds. Other work also suggested that SMBHs from stellar seeds do exist at high redshifts \citep{Fragione2023,Evans2023}, but are too faint to be currently observable. Therefore, in this study, we aim to reproduce the newly observed massive high redshift ($z\sim8-10$) AGN in cosmological zoom-in simulations by utilizing the DCBH seeding model instead. From the growth histories of the DCBH seeds, we specifically study the environmental conditions that allow for efficient SMBH growth to match observations.  

This paper is organized as follows. In Section \ref{sec:method}, we describe the subgrid models of galaxy formation and black hole evolution, including the new DCBH growth model employed here, as well as our simulation setup. In Section \ref{sec:results}, we show the different classes of DCBH and host halo evolution histories that we encounter in our simulations, test if such growth trajectories are realistic, and study the environmental conditions that produce efficient DCBH growth. In Section \ref{sec:observe} we present predicted spectra for our simulated AGN, and discuss their relation to the high-redshift sources currently observed with \textit{JWST}. We summarize our findings in Section \ref{sec:conclude}.

\section{Numerical Methodology} \label{sec:method}
We run cosmological zoom-in simulations of high-density regions to produce multiple DCBH seeds and study their growth. Previous studies have used cosmological simulations to study AGN growth and evolution as well \citep[e.g.][]{Somerville2015,Griffin2019,Jeon2023}. 
We use the \textsc{gizmo} code \citep{Hopkins2015} that inherits the gravity solver from the \textsc{gadget-2} framework \citep{Springel2005} and includes accurate Lagrangian solvers for hydrodynamics. We use the modified version of \textsc{gizmo} \citep{Liu2020,Jeon2023} with updated sub-grid models for star formation, stellar feedback, black hole formation, accretion, and feedback on top of the models for primordial chemistry, cooling, and metal enrichment in \citet{Jaacks2018,Jaacks2019}. We do not directly adapt subgrid models from other large simulation projects, such as FIRE-2 \citep{Hopkins2018}, EAGLE \citep{Schaye2015}, and AURIGA \citep{Grand2017}, as they have been generally calibrated/tested against lower redshift and local observations and are widely used to study such systems. However, our goal in this paper is to study SMBH formation at high redshifts within the first galaxies, and it has been shown with recent JWST observations that most pre-launch simulations, including those that have been very successful in explaining observations at lower redshifts, have underpredicted the observed number of galaxies at early times \citep{Finkelstein2023,Finkelstein2024,Austin2023}. Some limitations of the subgrid models calibrated against local observations may be due to simulation resolution, as targeted simulations performed at higher resolutions could reproduce the \textit{JWST} observations \citep[e.g.][]{Feldmann2024}. We therefore use the models developed in \citet{Jaacks2018,Jaacks2019,Liu2020} which have been specifically validated against high-resolution, ab-initio simulations of first galaxy formation, as well as observations at high redshifts. Because our subgrid models are thus not calibrated with observations across the full cosmic time, our simulation results are subject to significant uncertainties, emphasizing the exploratory nature of this study. Specifically, we use the Lagrangian meshless finite mass (MFM) hydro solver for our simulation runs. 

\subsection{Star formation and feedback}\label{star_formation_model}

We employ the stochastic star formation and feedback models developed in previous works \citep[see][]{Jaacks2018,Jaacks2019,Liu2020}. In those studies, the stellar models described below have been shown to be physically justified and numerically validated. Specifically, the Pop III models are calibrated to be consistent with extremely high-resolution zoom-in simulations \citep[e.g.][]{Johnson2007,Ritter2012,Ritter2015} and the second generation, metal enriched Population II (Pop II) stellar feedback modes to reproduce the observed cosmic star formation histories in $z\sim5-15$. We cannot resolve individual stars in the simulation, so each stellar particle represents a stellar population, characterized by Pop~III and Pop~II initial mass functions (IMFs) from \citet{Jaacks2018,Jaacks2019}. A gas particle is identified as an SF candidate when its hydrogen number density $n_{\rm H}>100$ cm$^{-3}$ and temperature $T\leq10^3$ K. For each SF candidate, the probability of spawning a stellar particle is 
\begin{equation}
    p_{\rm SF} = \frac{m_{\rm SF}}{m_*}[1-\exp(-\eta_*\Delta t/t_{{\rm ff},i})]\mbox{\ ,}
\end{equation}
where $m_{\rm SF}$ is the mass of the candidate gas particle, $m_*$ the mass of the stellar particle to be formed, $\eta_*$ the SF efficiency, $\Delta t$ the simulation timestep, and $t_{{\rm ff},i}= \sqrt{3\pi/(32G\rho_i)}$ the free-fall timescale of the gas particle with density $\rho_i$. As usual, $G$ is the gravitational constant. The simulation incorporates both the first generation, metal-free stars (Pop III), and the second generation, metal-enriched stars (Pop II). For Pop III stars, we set $\eta_*=0.05$ and for Pop II $\eta_* = 0.1$ \citep{Jaacks2019}. For both populations, we set $m_*\simeq600$ M$_\odot$ based on high-resolution Pop III star formation simulations \citep{Bromm2013,Stacy2016}, and further motivated by the timing of the global 21-cm absorption signal suggested by the Experiment to Detect the Global Epoch of Reionization Signature (EDGES) \citep{Schauer2019}. A random number $p$ is generated for the uniform distribution $[0,1]$ and a stellar particle is formed when $p<p_{\rm SF}$.

For stellar feedback, we include Lyman-Werner (LW) radiation, photoionziation heating, and supernova (SN) explosions \citep{Ciardi2005}. LW radiation is particularly crucial for this work as it creates environments that allow for DCBH formation -- destroying molecular hydrogen and slowing down gas cooling so that collapse is delayed until more massive halos $(>10^7$ M$_\odot$) emerge, which prevents fragmentation into less massive stars and produces the high inflow rates required to form massive black hole seeds. For LW radiation, we include two types:
A global LW background from the Pop III/II star formation rate density (SFRD) at a given time, and a local LW field in the neighborhood of a newly born star cluster for a given period \citep{Haiman1997,Johnson2007}. The total LW flux at time $t$ and position $\vec{x}$ is thus the sum of the global and local fields:
\begin{equation}
    J_{\rm LW}(t,\vec{x}) =  J_{\rm LW,global}(t)+ J_{\rm LW,local}(t,\vec{x})\mbox{\ .}
\end{equation}

We implement photoionization heating from stars similarly, with a global and a local component. The global heating is derived from a redshift-dependent photoionization rate. This rate is determined from the UV background produced by stars \citep{Faucher2009}. Local heating is applied to gas particles within the Str\"{o}mgren radius of active star particles on the fly. The total photoionization heating is the sum of the global and local heatings. We also include a model of self-shielding of gas particles against UV background photons based on equations 1 and 2 of \citet{Rahmati2014}.  

For SN feedback, we cannot resolve individual explosions \citep{Greif2007,Ritter2016}. Thus, we instead imprint the metal enrichment and thermal energy produced by a given SN event. After a typical stellar lifetime (3 Myr for Pop III and 10 Myr for Pop II) passes, the star particle ``dies". The total metals produced by the SN are distributed evenly to the gas particles inside the final radius of the expanded SN shell. This final shell radius depends on the total energy of SN explosions in the stellar population. For Pop~III stars, the SN energy and mass of heavy chemical elements produced to be distributed are calculated on the fly by counting progenitors sampled from the Pop III IMF, comprising of core-collapse and pair-instability SNe. For Pop II stars, we use IMF integrations to calculate the SN energy as $E_{\rm SN} \simeq 10^{52}$ erg $\times~m_\star/(10^3$ M$_\odot)$ and the produced metal yield as $M_Z = 0.016m_\star$, where $m_\star$ is the stellar population mass of the particle. We also model the thermal feedback from Pop~III SNe by imparting thermal energy to the nearby gas particles inside the Str\"{o}mgren radius - increasing the temperature by $2\times10^4$~K. In addition, the hydrogen in these gas particles is instantaneously ionized.

\begin{deluxetable*}{cccccccc}[htb!]
\tabletypesize{\footnotesize}
\tablenum{1}
\tablecolumns{1}
\tablecaption{Properties of the two mass definitions used to characterize the DCBH seed \label{bhmassdefinition}}
\tablewidth{0pt}
\tablehead{
\colhead{Definition} & \colhead{Mass at formation} & \colhead{Mass accretion method} &\colhead{Purpose in simulation} &\colhead{Considered as the physical mass}\\ \colhead{} & \colhead{} & \colhead{} & 
\colhead{} 
}
\startdata
$M_{\rm max,BH}$ &10$M_{\rm gas}$ &Eddington accretion & \makecell{Define the most ideal  \\ black hole growth}  & No \\
$M_{\rm BH}$ &$M_{\rm gas}$ & \makecell{Swallow nearby gas \\ to try match $M_{\rm max,BH}$} & \makecell{Follow realistic \\ black hole growth} & Yes \\
\enddata
\centering
\end{deluxetable*}

Finally, we account for mechanical feedback through Pop II star SN-driven winds \citep{Springel2003}. The probability for a Pop~II SF candidate to be launched as a wind particle before being able to spawn a stellar population is 
\begin{equation}
    p_w = 1-\exp\left(-\eta_{w,{\rm SF}}\frac{m_\star}{m_{\rm SF}} \right) \mbox{\ ,}
\end{equation}
where $\eta_{w,{\rm SF}}=2$ is the wind-loading factor. A random number $p$ is generated and the wind particle is launched when $p<p_w$, similar to the SF routine. After being launched, the gas particle receives a kick of $\simeq170$ km s$^{-1}$ in a random direction. This velocity is derived from a SN wind model dependent on the energy of the SN and the couplings between the SN energy, heated gas, and wind \citep{Springel2003}. The ejected gas particle is decoupled from hydrodynamics and cannot be considered as a SF candidate before a time of $0.1H(z)^{-1}$ has passed, or its (hydrogen number) density has dropped below $10\ \rm cm^{-3}$, where $H(z)^{-1}$ is the local Hubble time. We do not apply this model to Pop~III stars, as the full impact of Pop III SNe has been captured by the photoionization heating and thermal energy injection. The full details and equations for the feedback physics can be found in \citet{Liu2020}.

\subsection{Black hole seeding}

We incorporate both Pop III stellar remnant and DCBH seeds in our simulations. For Pop III seeds we specifically use the more massive seed model of \citet{Liu2020}. Accordingly, once a star particle reaches the end of its life at 3~Myr, we assume that stars contained in the particle with $m_{\ast}$ in the range $40\leq m_{\ast}/\text{M}_\odot \leq 140$, sampled from the IMF, will merge and form a single black hole. Their total stellar mass is converted to the black hole mass and the stellar particle is turned into a black hole sink particle. However, as stellar seed black holes are generally much lower in mass compared to DCBH seeds and do not grow efficiently under the standard models \citep{Jeon2023}, they do not affect our results significantly.

For DCBH seeding, we impose conditions on the gas density, temperature, and chemistry to model the dense, hot, and metal-poor environment in which DCBHs can form \citep[e.g.,][]{Bromm2003,Ardaneh2018,Wise2019,Chon2021}. For a gas particle to become a DCBH, we require that its (hydrogen number) density reaches $n_{\rm DCBH} \geq 2\times10^3$ cm$^{-3}$, and that it has a temperature of $7000<T/\text{K}<10^4$, a metallicity of $Z<Z_{\rm DCBH} = 2\times10^{-4}$ Z$_\odot$, and H$_2$ abundance of $x_{\rm H_2}<10^{-6}$, following \citet{Liu2020}. Unlike \citet{Liu2020} however, we also require that the gas particles in the neighborhood of the DCBH candidate (within the softening length) also satisfy the temperature, metallicity, and H$_2$ abundance criteria, and that the DCBH candidate is the densest gas particle among its neighbors. Therefore, a DCBH sink particle will form from the densest particle in a region of gas that meets the chemical and temperature criteria. We tighten the conditions for DCBH formation because the zoom-in simulation we run targets denser and more extreme regions than that of \citet{Liu2020,Jeon2023}, so that considering only the individual gas particle conditions may overproduce DCBH seeds. To reiterate, it is expected that for DCBH formation, the environment will need to exhibit (chemically) near-primordial conditions \citep{Natarajan2017}.

When a gas particle meets the criteria above and is converted to a DCBH, we use two different mass values to characterize the sink particle. Table~\ref{bhmassdefinition} summarizes the two mass definitions. The first is the ``maximal black hole mass'', $M_{\rm max,BH}$, set at DCBH formation to 10 times that of the original gas particle mass $M_{\rm gas}$. This mass represents the maximal mass the black hole may grow through accretion at the Eddington rate (Section~\ref{sec:bhacc}). The second is the actual black hole mass, $M_{\rm BH}$, set equal to $M_{\rm gas}$ at DCBH formation. $M_{\rm BH}$ represents the total sink particle mass, including the neighboring stars, their remnants, and any gas bound to the black hole\footnote{In terms of terminology, we note that $M_{\rm max,BH}$ corresponds to $M_{\rm BH}$ in \citet{Liu2020,Jeon2023}, whereas $M_{\rm BH}$ corresponds to the `dynamical mass' $M_{\rm dyn}$ in \citet{Liu2020,Jeon2023}}. We set the initial $M_{\rm max,BH}$ to 10 times the gas particle mass, because DCBH seeds are expected to have masses of $\sim10^5-10^6$ M$_\odot$ \citep{Pacucci2017,Becerra2018a,Becerra2018b,Basu2019}, while our gas particle masses are $\sim10^4-10^5$ M$_\odot$ (see Section \ref{sec:simset}). However, manually increasing a particle mass in the simulation as such would violate mass conservation. Therefore, for this work, we use the second mass definition $M_{\rm BH}$ to represent the actual mass of the black hole (and its nearby components) and use $M_{\rm max,BH}$ only to calculate the accretion rate (see Section \ref{sec:bhacc}). To obey mass conservation but also to produce massive seeds, we allow the black hole seeds to stochastically swallow nearby gas particles \citep{Springel2005_2,Liu2020}. For $M_{\rm max,BH}\geq10^4$ M$_\odot$, which all our DCBH seeds are initialized to, $M_{\rm BH}$ is only updated when the sink particle can swallow a gas particle (Section~\ref{sec:bhacc}). Thus, when $M_{\rm max,BH}$ is set to be larger than $M_{\rm BH}=M_{\rm gas}$ at DCBH formation, the black hole particle will try to swallow nearby gas particles to match the two masses and will maintain mass conservation for $M_{\rm BH}$. With our criteria for DCBH formation, the seeds generally form in high-density environments so that the seed can immediately swallow the nearby gas particles after formation to reach $M_{\rm BH}\sim M_{\rm max,BH}$. 

\subsection{Black hole accretion and growth} \label{sec:bhacc}

\begin{deluxetable*}{cccccccc}[htb!]
\tabletypesize{\footnotesize}
\tablenum{2}
\tablecolumns{1}
\tablecaption{Summary of the zoom-in simulation parameters and the parent simulations used to define the zoom-in regions. Length units are in comoving coordinates. \label{simtable}}
\tablewidth{0pt}
\tablehead{
\colhead{Run} & \colhead{\makecell{Parent linear \\ box size}} & \colhead{\makecell{Parent \\ particle number}} &
\colhead{\makecell{Parent \\ particle type}} & \colhead{\makecell{Linear zoom-in \\ box size}} & \colhead{\makecell{Zoom-in \\ DM mass}} &\colhead{\makecell{Zoom-in \\ gas mass}} & \colhead{\makecell{Zoom-in average (minimum)\\ gravitational softening length}} \\ \colhead{} & \colhead{$h^{-1}$ Mpc} & \colhead{} & 
\colhead{} & \colhead{$h^{-1}$ Mpc} & \colhead{M$_\odot$} &\colhead{M$_\odot$} & \colhead{$h^{-1}$ kpc} 
}
\startdata
\textsc{zoom12}  &$12$ &$2\times128^3$  &DM and baryons  &$1.44$ &$1.4\times10^6$& $2.5\times10^5$& 2.3 (0.2) \\
\textsc{zoom16} &$16$  &$256^3$ &DM only &$2$ & $4.2\times10^5$& $7.5\times10^4$& 1.5 (0.2) \\
\textsc{zoom12\_HR}  &$12$ &$2\times128^3$  &DM and baryons  &$1.44$ &$1.8\times10^5$& $3.1\times10^4$& 1.2 (0.2)  \\
\enddata
\centering
\end{deluxetable*}

For stellar seed black holes, we apply the original Bondi-Hoyle equation \citep{Bondi1944} used in \citet{Jeon2023} to determine black hole accretion. However, for DCBH seeds, we enforce $M_{\rm max,BH}$ to grow at the Eddington rate given by
\begin{equation}
\dot{M}_{\rm Edd} = 2.7\times10^{-3}\left(\frac{M_{\rm max, BH}}{10^5~\text{M}_\odot}\right)\left(\frac{\epsilon_0}{0.1}\right)^{-1}\rm~M_\odot~\text{yr}^{-1}\mbox{\ ,}
    \label{edd}
\end{equation}
where $\epsilon_0=0.125$ \citep{Negri2017}. As this exponential growth is forced and so most likely would violate mass conservation, we do not consider $M_{\rm max,BH}$ as the actual black hole mass as in previous works \citep[e.g.][]{Liu2020,Jeon2023}. We only use $M_{\rm max,BH}$ associated with the black hole sink particle to calculate $\dot{M}_{\rm Edd}$, and instead use $M_{\rm BH}$ to track the actual growth of the black hole mass. $M_{\rm BH}$ tries to match the Eddington growth by swallowing nearby gas particles in the black hole kernel and only updates its mass if it can accrete gas particles. The kernel size is adaptive to include 64 gas particles nearest to the black hole with an upper limit of 1 comoving kpc/$h$. This upper limit was set as the approximate lower limit of a halo radius and varying this limit did not affect the DCBH growth histories significantly.

To model the gas swallow process, we adapt the probabilistic procedure from \citet{Springel2005_2}. The probability that a gas particle in the black hole kernel will be swallowed (accreted) is evaluated as
\begin{equation}
    p_{\rm acc} = \frac{w\dot{M}_{BH}\Delta t}{\rho}\mbox{\ ,}
\end{equation}
where $w$ is the kernel weight of the gas particle relative to the black hole, $\dot{M}_{\rm BH}$ the black hole accretion rate, $\Delta t$ the time step, and $\rho$ the gas density estimated at the black hole position. A random number $p$ is drawn and if $p<p_{\rm acc}$ the gas particle will be swallowed. Thus, even if $M_{\rm max,BH}$ grows at the Eddington rate with $\dot{M}_{\rm max,BH}=\dot{M}_{\rm Edd}$, if the environment near the DCBH has a low probability to swallow gas, whether it be due to the lack of gas inflow, high gas temperature, or the dark matter halo gas mass limit, $M_{\rm BH}$ will not be able to increase.

We apply such a method for DCBH seed growth as previous studies have found that DCBHs with masses greater than $10^5$ M$_\odot$ could undergo efficient growth with Eddington ratios as large as 100 starting at formation \citep{LiHernquist2007,Pacucci2017,Basu2019}. However, our simulation cannot resolve the physics determining the accretion rate in the close vicinity of the black hole, and we have to revert to  sub-grid models. This limited resolution can lead to underestimating the black hole accretion rate in cosmological simulations \citep{Dimatteo2012,Schaye2015,Trinca2022,Jeon2023}. Moreover, the choice of the accretion subgrid model can lead to different black hole growth histories in the same simulation \citep{Soliman2023}. Therefore, rather than choosing a specific subgrid model for DCBH accretion, we impose that DCBHs will be able to accrete efficiently as long as the environmental conditions allow them to enact this gas-swallowing procedure for $M_{\rm BH}$. The swallowing of gas particles also limits SF in the vicinity of the DCBH, consistent with previous work that showed suppressed SF in the vicinity of an AGN \citep{Hopkins2023}. In reality, galactic cooling flows \citep{Fabian2023} or instabilities caused by magnetic fields and non-uniform densities \citep{Jiang2014,Jiang2019,Davis2020} may lead to such efficient accretion of nearby gas. We are thus able to determine what kinds of cosmological environments allow for efficient black hole growth to reproduce the high redshift AGN observations.

We found that the black hole thermal feedback, as modeled in \citet{Liu2020,Jeon2023}, is negligible for the overall evolution of the cosmological environment in this work. Black hole thermal feedback injects energy into the gas particles in the black hole kernel. The total energy to be distributed during a given time step $\delta t$ is $\delta E = \epsilon_r L_{\rm BH}\delta t$, where $\epsilon_r=0.02$ is
the radiation-thermal coupling efficiency \citep{Tremmel2017}, and $L_{\rm BH}$ the black hole luminosity
\begin{equation}
    L_{\rm BH} = \epsilon_{\rm EM}\dot{M}_{\rm acc}c^2\mbox{\ .}
\end{equation}
Here, the radiative efficiency $\epsilon_{\rm EM}$ is defined as
\begin{equation}
    \epsilon_{\rm EM} = \frac{\epsilon_0A\eta}{1+A\eta},~\eta\equiv\dot{M}_{\rm acc}/\dot{M}_{\rm Edd}\mbox{\ ,}
\end{equation}
with $\epsilon_0=0.125$ and $A=100$ \citep{Negri2017}. The thermal feedback model directly only affects the gas inside the hydro kernel of the black hole particle. As there are only a small number of heavy seeds in the simulation box that accrete efficiently, relatively few gas particles are affected by BH thermal feedback and do not contribute significantly to the overall cosmological evolution. Furthermore, the black hole kernel is the same region that the gas particles to be swallowed are drawn from. Therefore, under our assumption of efficient accretion using the gas swallow subgrid model, we found that the gas particles inside the black hole kernel will generally be swallowed eventually, after enough time passes for the gas to be transported to near the black hole, even if they are affected by thermal feedback. The thermal feedback can actually self-regulate black hole accretion, as the probability of swallowing the heated gas particles that are driven away from the black hole will be lower. Thus, a high accretion rate will lead to hotter gas and lower probability of swallowing/accretion rates, and low accretion to colder gas and higher accretion rates. To test this effect, we have performed a simulation run without the thermal feedback being present, and found that on average the black holes grew to higher masses (by a factor of $\sim2$) without feedback. We also do not apply the black hole mechanical feedback model from swallowing of gas particles used previously \citep{Liu2020,Jeon2023} as we are looking for optimal cosmological conditions for black hole growth and want to obtain optimistic estimates of black hole masses. The model also overestimates the wind strength from massive black hole particles. This was not a problem in previous works as no massive black holes formed then.

\subsection{Simulation setup}\label{sec:simset}

We use two zoom-in regions identified in two parent simulations. Regions of different sizes at different resolutions are used to test the consistency of the gas swallow growth model. For all cases, we use \textit{Planck} cosmological parameters \citep{Planck2016}: $\Omega_m = 0.315$, $\Omega_b = 0.048$, $\sigma_8 = 0.829$, $n_s = 0.966$, and $h = 0.6774$ and initial conditions are generated with the \textsc{MUSIC} code \citep{Hahn2011} at $z=99$. 

The first parent simulation run includes both dark matter (DM) and baryons in a box of comoving linear size of $L_C \sim12\ h^{-1}$ Mpc with $2\times128^3$ gas and DM particles. We run the simulation until $z=5.2$ and identify dark matter halos in post-processing with \textsc{Rockstar} \citep{Behroozi2013}. We identify the halo (with mass $\sim10^{11}$ M$_\odot$) that hosts the most massive SMBH ($\sim2\times10^8$ M$_\odot$) by $z=5.2$. We note that the SMBH mass may be overestimated in this case due to the lower mass resolution of the parent simulation. From the initial conditions at $z=99$, we identify a box of comoving linear length $L_C \sim1.44\ h^{-1}$ Mpc that contains all the particles of the chosen $z=5.2$ halo. We rerun the simulation using this box as the zoom-in region (\textsc{zoom12}) with increased length and mass resolution by a factor of $2^2$ and $8^2$, respectively, compared to the parent simulation. We use an adaptive gravitational softening length for particles based on their kernel size with a minimum length of 0.2 $h^{-1}$\,kpc. The effective (average) softening length for the zoom-in run is $\varepsilon\sim 0.1 L_C/N^{1/3}\sim$ 2.3 kpc $h^{-1}$, where $N$ is the number of particles in the zoom-in box.

The second parent simulation run includes only DM particles in a box of comoving linear size of $L_C \sim16\ h^{-1}$ Mpc with $256^3$ particles. We run the simulation until $z=7$, identify dark matter halos in post-processing, and find the most massive dark matter halo ($\sim10^{11}$ M$_\odot$). From the initial condition, we choose as the zoom-in region (\textsc{zoom16}) a box with comoving length $L_C \sim2\ h^{-1}$ Mpc that contains all the particles of the $z=7$ halo. This region is rerun with baryons and higher resolution by a factor of  $2^2$ (length) and $8^2$ (mass). The effective softening length for this zoom-in run is 1.5 $h^{-1}$\,kpc. 

\begin{figure*}[htb!]
\gridline{
 \fig{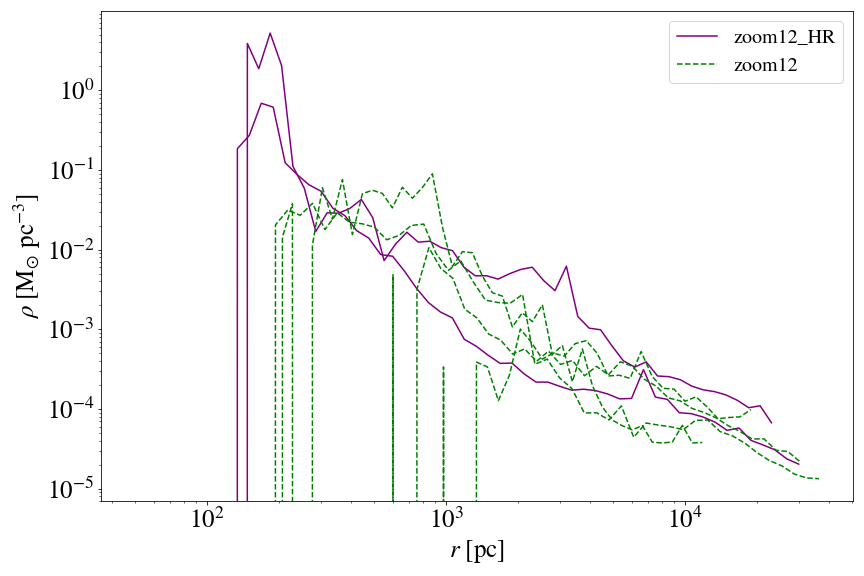}{0.5\textwidth}{}
 \fig{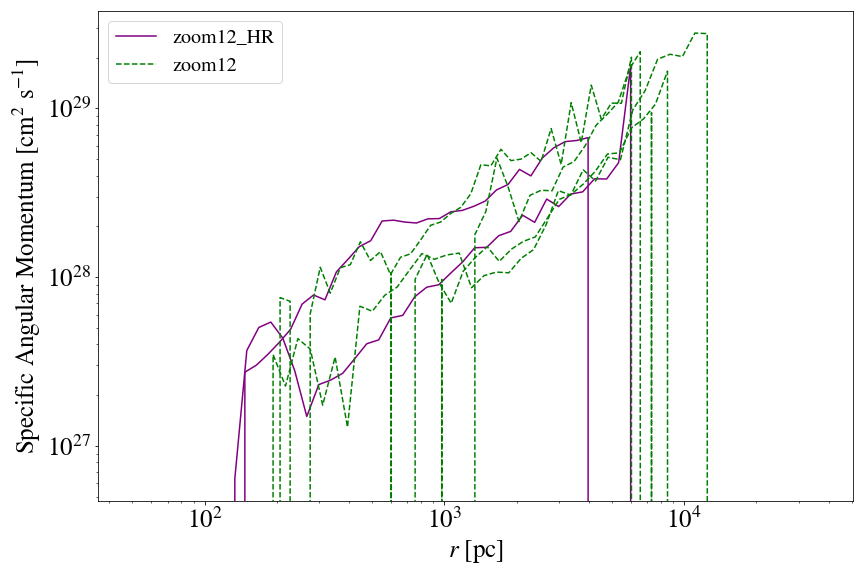}{0.5\textwidth}{}
 }  \caption{Profile plots of the gas density ({\it left}) and specific angular momentum ({\it right}) against distance from the halo center ($r$), for the halos hosting heavy seeds at $z=9.4$ within the \textsc{zoom12} (lower resolution) and \textsc{zoom12\_HR } (higher resolution) runs. The density and angular momentum evolution of the halos are generally consistent regardless of the simulation resolution, confirming that our chosen resolution for \textsc{zoom12} should be adequate to study high-redshift systems and heavy black hole seed evolution. }
    \label{fig:resolution}
\end{figure*}

Lastly, we run a higher resolution version of \textsc{zoom12} with $\sim10$ times more particles in the zoom-in region, \textsc{zoom12\_HR}, to test the consistency of the simulation results under different resolution, albeit to not as low a redshift due to limited computational resources. We compare the halo gas density and angular momentum profiles for halos hosting DCBH seeds at $z=9.4$ in Fig.~\ref{fig:resolution} and find that they are consistent between \textsc{zoom12} and \textsc{zoom12\_HR}, thus validating the numerical choices employed in the \textsc{zoom12} simulation, used for our main analysis (see Sections~\ref{sec:results} and \ref{sec:observe}). In Table \ref{simtable}, we summarize the simulation parameters.

After running the zoom-in simulations, we post-process the output with \textsc{yt} \citep{Turk2011} and examine DCBH formation and evolution throughout the runs.

\section{Results\label{sec:results}}
From the simulation output, we analyze the formation and growth of the DCBH seeds. We also examine the DCBH host halo properties identified by \textsc{rockstar}. Focusing on the mass evolution of the stars in the halo and the DCBH seed, we find a variety of growth histories a halo-AGN system can have. We then study the evolution of the environment near the host halos and AGN to determine the conditions for efficient growth. For \textsc{zoom12}, 12 DCBH seeds formed and for \textsc{zoom16}, 7 seeds formed. The lower number of seeds for \textsc{zoom16} may be due to the fact that it was not run to as low a redshift as \textsc{zoom12} because of computational limits.

\subsection{AGN growth histories \label{sec:growth}}

\begin{figure*}[htb!]
    \centering
    \includegraphics[width=\textwidth]{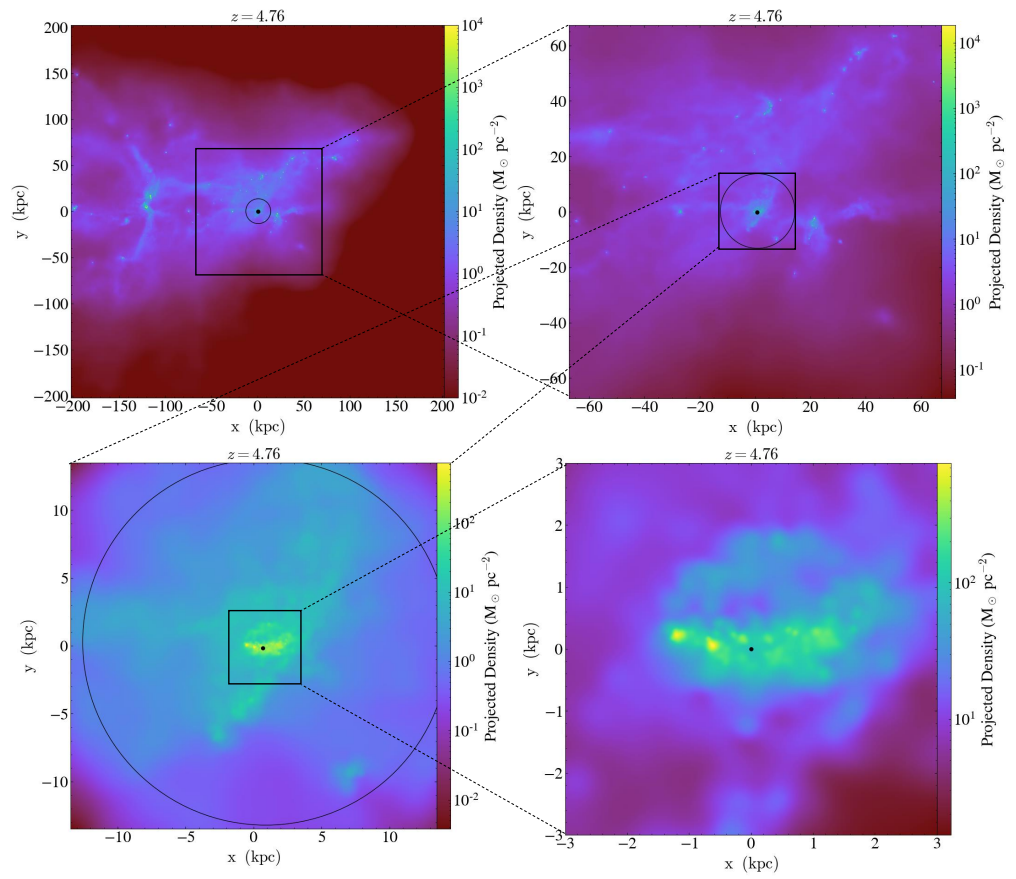}
\caption{DCBH seed environment. We show gas densities, in projection, around a growing DCBH by the end of the \textsc{zoom12} simulation run at different scales. Distances are in physical units. The black circle ({\it left-upper panel}) marks the virial radius of the halo and the black dot the position of the DCBH, with its size not to scale. The DCBH seed resides in an environment with a higher gas density than average, which aids the SMBH accretion and growth.
\label{fig:projectionplot}}
\end{figure*}

\begin{figure*}[htb!]
\gridline{
 \fig{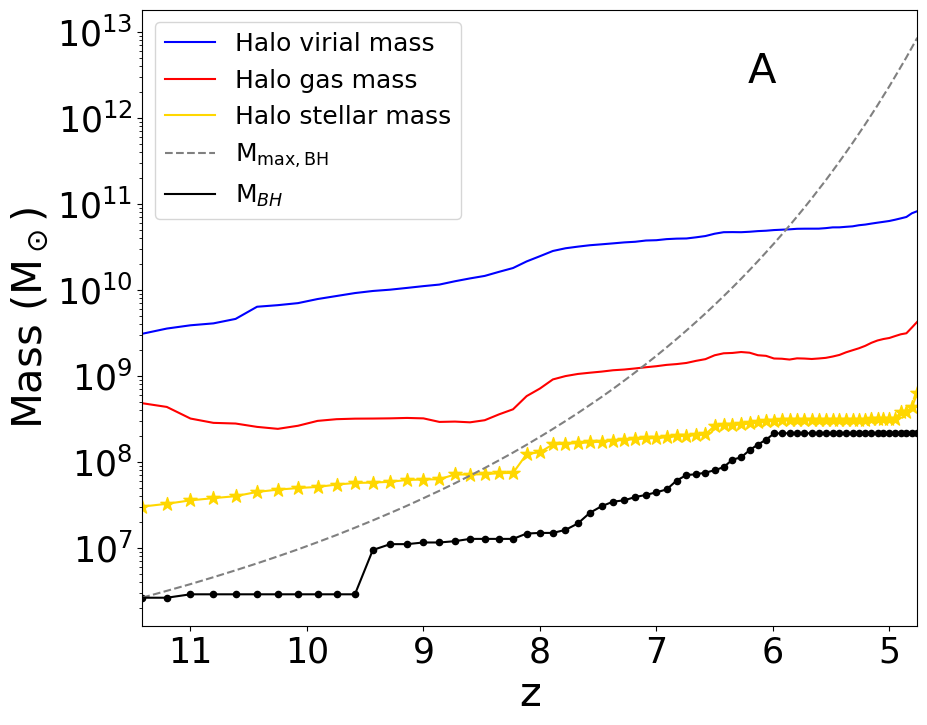}{0.5\textwidth}{}
 \fig{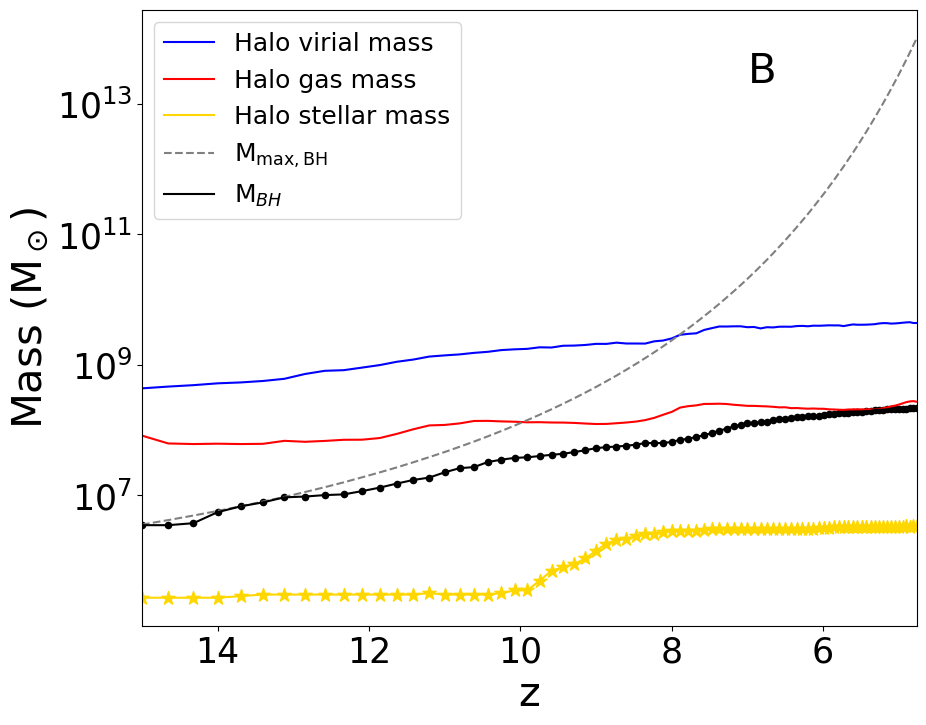}{0.5\textwidth}{}
}
\vspace{-0.50cm}
\gridline{
 \fig{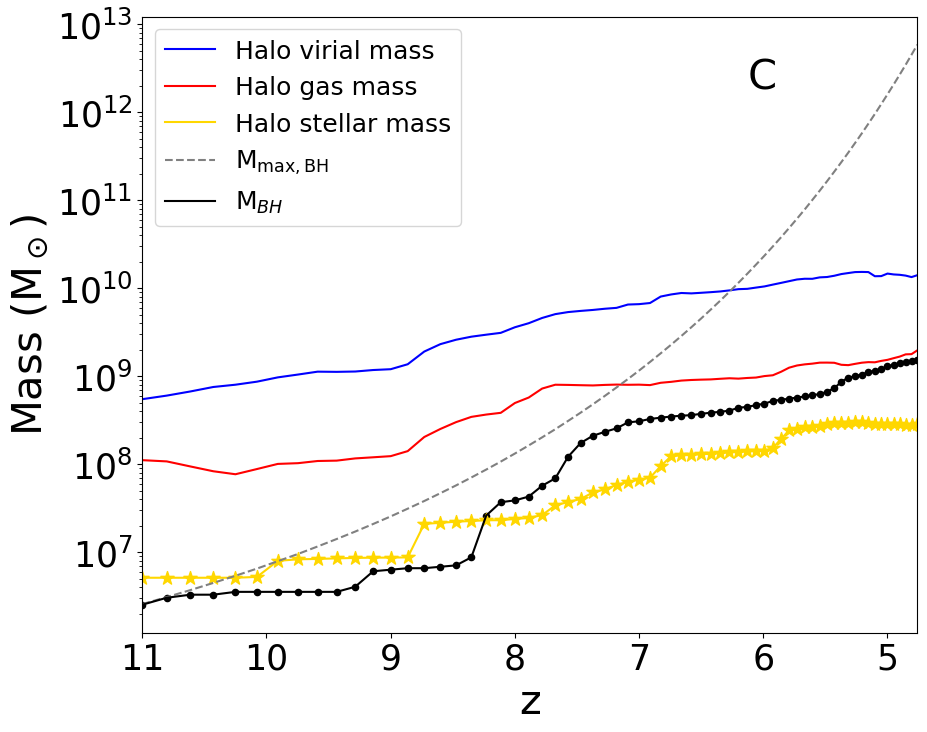}{0.5\textwidth}{}
 \fig{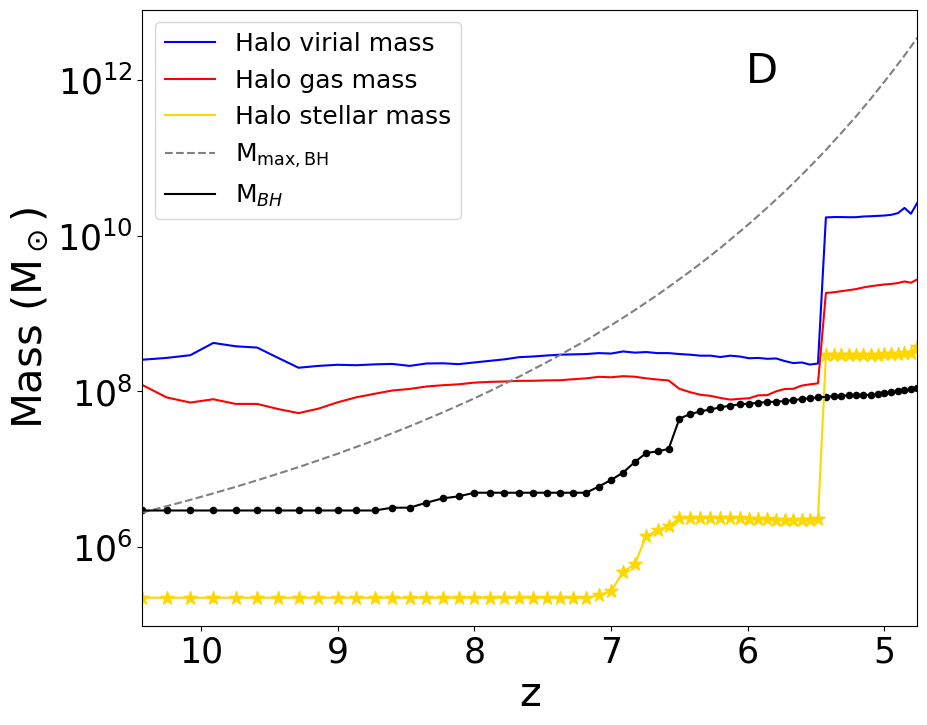}{0.5\textwidth}{}
}
\caption{Examples of different mass accretion histories of DCBH seeds and host halos. We plot both the actual black hole mass growth ($M_{\rm BH}$) and the black hole mass growing at the Eddington rate ($M_{\rm max,BH}$). The DCBHs formed stochastically during the \textsc{zoom12} run. We see various relationships between the DCBH and halo stellar mass growths: A, B show cases where one mass is always above the other, whereas C, D illustrate cases where there is a turnover in the masses. Panel~D specifically shows how the stellar mass is overtaking the AGN mass through a merger with a different halo at $z\sim5.5$. As can be seen, DCBH mass growth follows different trajectories, each experiencing Eddington-limited phases, interspersed with episodes of slower growth. The wide variety of assembly histories for DCBHs under the same accretion model suggests that environmental factors are significant in determining how efficient AGN growth progresses. 
\label{fig:massplot}}
\end{figure*}

We show the gas environment around one DCBH seed by the end of the \textsc{zoom12} simulation run in Fig.~\ref{fig:projectionplot}. The environment that the DCBH seed resides in has higher gas density than average with the plot showing the seed located in a high-density node. Such conditions provide fuel to the DCBH seed and facilitate its accretion and growth.  We further illustrate in Fig.~\ref{fig:massplot} the possible AGN and host halo growth histories from our \textsc{zoom12} run. Various classes of growth histories exist and we show examples categorized by the relationship between the AGN and the host halo stellar mass: The stellar mass can be always dominant over the AGN mass (Fig.~\ref{fig:massplot}A), the AGN mass always dominate over the stellar one (Fig.~\ref{fig:massplot}B), or one mass component might overtake the other due to more efficient growth (Fig.~\ref{fig:massplot}C) or mergers (Fig.~\ref{fig:massplot}D). We note that Case~C corresponds to the DCBH seed shown in Fig.~\ref{fig:projectionplot}.

\begin{figure*}[htb!]
    \centering
    \includegraphics[width=\textwidth]{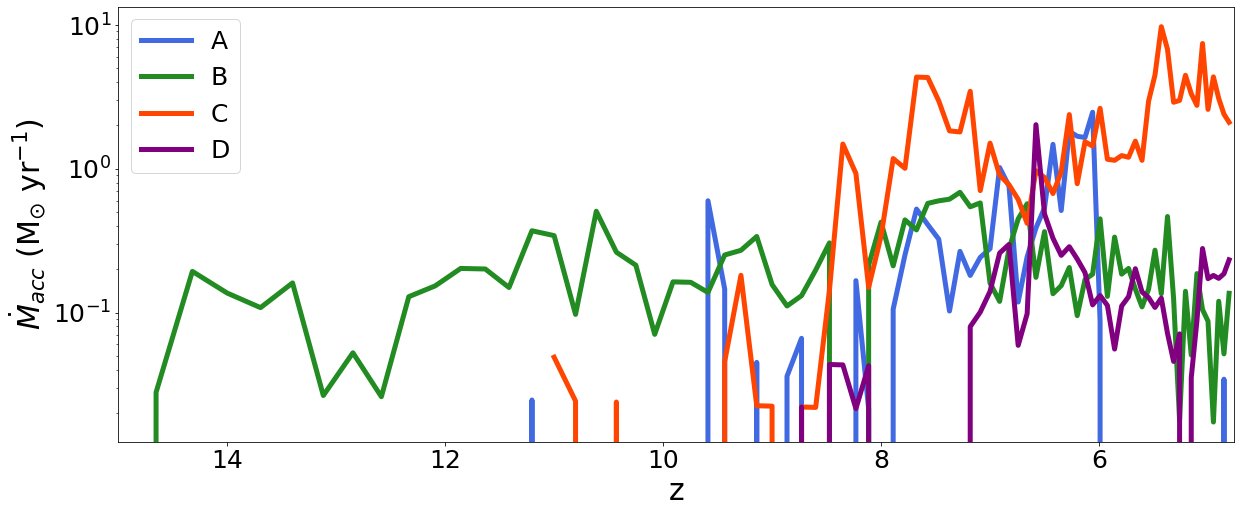}
    \caption{Mass accretion rates for DCBH seeds, as shown in Fig.~\ref{fig:massplot}, across redshifts. If the rate is not plotted for a given redshift, it indicates that the accretion was insignificant in that time period. Comparing with Fig.~\ref{fig:massplot}, one can see that for DCBH seeds with more steady accretion, cases B and C, the AGN mass dominates over the stellar one. In contrast, cases A and D show more episodic mass evolution and the stellar mass dominates over the AGN mass by the end of the simulation. As each seed resides in halos of similar virial and gas masses, differences in gas and stellar distributions near the seed black hole are likely responsible for such variations in mass accretion trends.}
    \label{fig:dcbhaccretion}
\end{figure*}

We further show the corresponding mass accretion rates for the seeds in Fig.~\ref{fig:dcbhaccretion}. The seeds within halos dominated by the AGN mass by the end of the simulation, B and C, experience more steady growth across redshifts, while the seeds where the stellar mass dominates by the end, A and D, exhibit intermittent accretion with periods of negligible accretion. Fig.~\ref{fig:massplot} indicates that the black holes reside in halos of similar virial ($\sim$10$^{9-10}$\,M$_{\odot}$) and gas masses ($\sim$20\% of virial). Seed~A actually resides in a halo with a slightly higher gas mass than the others, yet it experiences intermittent growth. Therefore, the resulting accretion does not only depend on the available amount of gas and stars in the host halo, but also on their distribution and properties near the black hole seed (see Section~\ref{conditions}).

\subsection{Testing the gas swallow model}\label{sec:justify}

We now argue that the high, near-Eddington, accretion rates enforced through the gas swallowing implementation can be physically justified. To do so, we consider different accretion regimes in post-processing: hot accretion, where thermal gas pressure is important and angular momentum is negligible, and cold accretion, where rotational motion of the accretion disk dominates over thermal pressure, such that $v_{\rm rot}>c_s$ where $v_{\rm rot}$ is the disk rotational velocity and $c_s$ the gas sound speed. We further model the hot accretion regime with the classical Bondi-Hoyle prescription, and the cold accretion case with a thin-disk model. 

\begin{figure*}[htb!]

\gridline{
 \fig{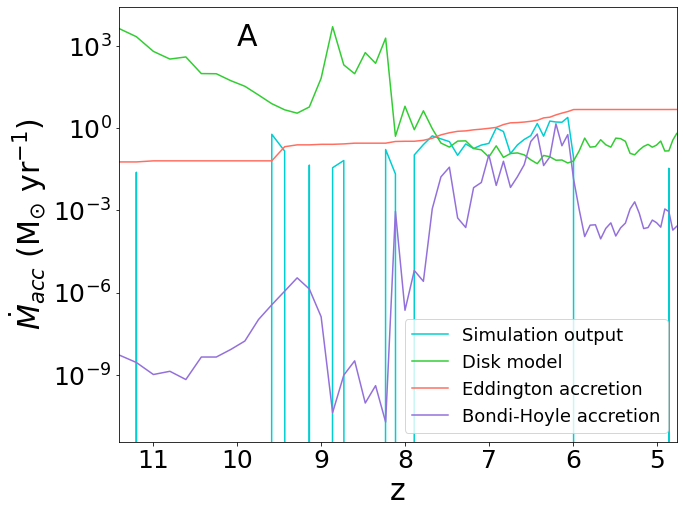}{0.5\textwidth}{}
 \fig{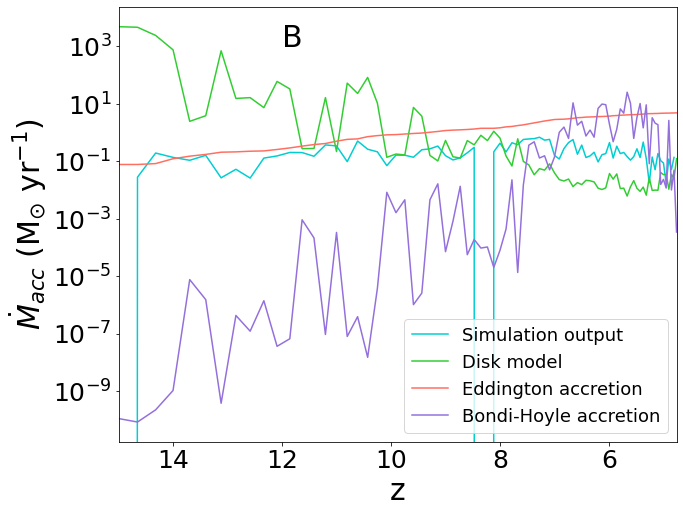}{0.5\textwidth}{}
}
\vspace{-0.50cm}
\gridline{
 \fig{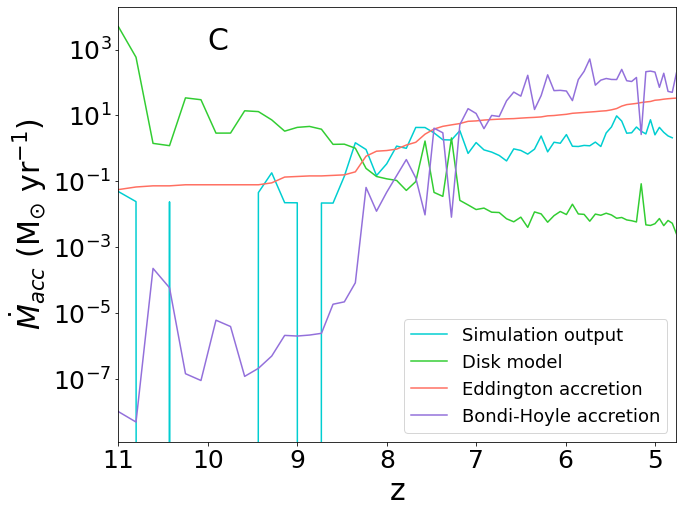}{0.5\textwidth}{}
 \fig{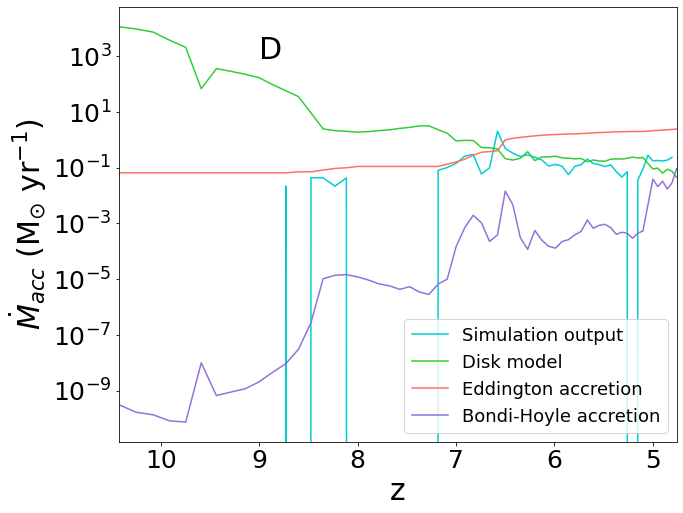}{0.5\textwidth}{}
}
\caption{Comparison of the accretion rates of the DCBH seeds in Fig.~\ref{fig:massplot} from the simulated gas swallow model, the accretion disk model (Equ.~\ref{diskequation}), the Eddington limit (Equ.~\ref{edd}), and the Bondi-Hoyle model (Equ.~\ref{bondi}). We find that the disk accretion model predicts higher accretion at early times, matches the other accretion rates at intermediate times, and exhibits less vigorous accretion at later times. This reflects the properties of the model, as at early times, high gas densities in the disk are predicted, while as time passes, the gas density and the accretion through the disk decreases. As can be seen, the gas swallow model in our simulations yields accretion rates that are largely similar to the Eddington, Bondi-Hoyle, and $\alpha$-disk values, at least for select periods of time.  
\label{fig:diskmodel}}
\end{figure*}

Specifically, the Bondi-Hoyle rate is given by
\begin{equation}
\dot{M}_{\rm Bondi} = \frac{4\pi(GM_{\rm BH})^2\rho_g}{c_s^3}   \mbox{\ ,}
\label{bondi}
\end{equation}
where $\rho_g$ is the average density of the gas particles in the black hole kernel, and $c_s$ the average sound speed for the same gas particles. Unlike accretion from a uniform gas cloud with zero angular momentum, as assumed in the Bondi-Hoyle model, SMBH accretion likely occurs through cold, thin accretion disks with viscosity and stress to enable mass accretion \citep{Jiang2019_2,Davis2020,Liu2022}. Although the accretion disk cannot be resolved in our simulations, so that we use the heuristic gas swallowing model instead, we can still estimate the expected disk-driven accretion rate, as follows

\begin{equation}
    \dot{M}_{\rm disk} = 3\pi\Sigma\alpha H_{\rm disk}c_s\mbox{\ ,}
    \label{diskequation}
\end{equation}
where $\Sigma$ is the gas surface density, $\alpha$ the viscosity parameter in the \citet{Shakura1973} thin-disk model, $H_{\rm disk}$ the scale-height of the disk, and $c_s$ the sound speed of the gas in the disk \citep[e.g.,][for a simplified derivation]{Bromm2013_2}. As the gas particles in the DCBH kernel will be swallowed eventually for the enforced Eddington accretion approach, we use their properties in estimating the accretion disk parameters. For the gas surface density, we divide the average mass of the gas particles within the kernel by the disk area with radius
\begin{equation}
    r_{\rm disk} = 10^4\frac{6GM_{\rm BH}}{c^2}\mbox{\ ,}
\end{equation}
where $c$ is the speed of light \citep{Jeon2014}. We here effectively assume that as the sink particle will swallow one gas particle at a time, the average gas particle mass will be uniformly distributed in an accretion disk. We choose $\alpha =1$, assuming that gravitational torques transport the angular momentum of the gas in the disk \citep{Bromm2013_2}. For the disk scale height, we use 
\begin{equation}
    H_{\rm disk}(r) = c_s\left(\frac{GM_{\rm BH}}{r^3}\right)^{-1/2}
    \mbox{\ ,}
\end{equation}
where the factor in the parentheses corresponds to the angular velocity of the disk at radius $r$ \citep{Shakura1973}. We evaluate the average value of $H_{\rm disk}(r)$ between $r_{\rm disk}$ and $r_{0} = 6GM_{\rm BH}/c^2$, the radius of the innermost stable orbit. The above equations can be combined to express $\dot{M}_{\rm disk}$ in terms of the gas and black hole properties as
\begin{equation}
    \dot{M}_{\rm disk} \propto \frac{M_{\rm gas}\alpha c_s^2}{M_{\rm BH}} \mbox{\ ,}
    \label{diskaccretion_prop}
\end{equation}

where $M_{\rm gas}$ is the average gas mass near the black hole. The accretion rate is inversely proportional to the black hole mass and proportional to gas mass, so to maintain high accretion, a continuously growing gas supply is needed to fuel a more massive accretion disk. 

We compare our two illustrative, physically motivated accretion rates calculated in post-processing, for the Bondi-Hoyle and thin disk models, with the rate obtained within our numerical gas-swallowing prescription in simulations, as well as the post-processed Eddington value in Fig.~\ref{fig:diskmodel}. The rate predicted by the disk model exceeds all others at early times, but declines at later times, reflecting the $M_{\rm BH}^{-1}$ dependence of the disk accretion rate in Equ.~(\ref{diskaccretion_prop}), while at the same time exhausting the available gas supply. In contrast, the Bondi-Hoyle model tends to predict low accretion rates at early times, but increased rates at late times, compared to the gas swallow model. As the BH mass increases with time, such behavior reflects the $M_{\rm BH}^2$ dependence of the Bondi-Hoyle accretion rate, see Equ.~(\ref{bondi}). Thus, even with our simplified thin-disk and spherical Bondi-Hoyle accretion models, we are able to produce accretion rates that are similar to the simulation output, and in turn the Eddington limit, at intermediate redshifts, when the high-redshift AGN have been observed ($z\sim7-9$). Even outside this range, the accretion disk and Bondi-Hoyle models generally do not yield extremely different values compared to the gas swallowing model. Thus, the growth and accretion rates of the DCBHs in our simulations are not unrealistic, and can be reproduced to first order with idealized, physically-motivated models.

\begin{figure}[htb!]
    \centering
    \includegraphics[width=0.5\textwidth]{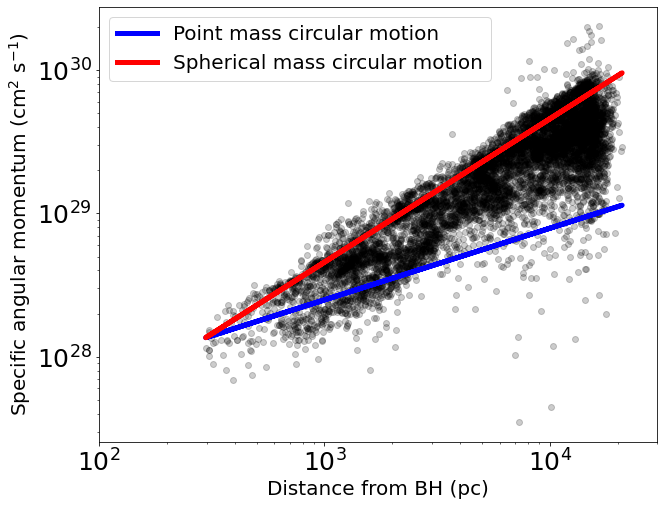}
    \caption{Specific angular momentum of the gas particles near the Case C DCBH seed in the simulation at $z=4.76$. We compare against analytical models that assume centrifugally-supported (circular) motion around a central mass. The first model assumes a point mass with the heavy seed at the center, and the second with the total mass enclosed within a given radius, normalized to match the point source model at the gas particle closest to the heavy seed. The gas particle angular momentum does not show rapid decrease as it gets closer to the black hole, and generally agrees with the analytical models.}
    \label{fig:haloang}
\end{figure}
\begin{figure}[htb!]
    \centering
    \includegraphics[width=0.5\textwidth]{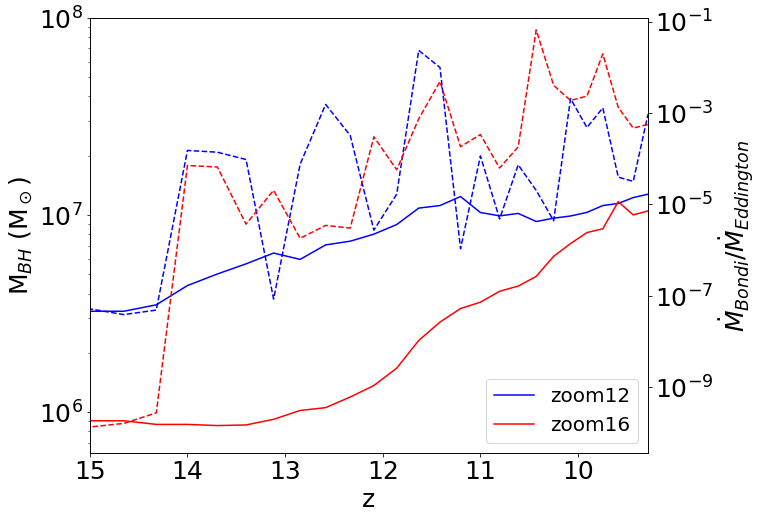}
    \caption{Dependence on cosmic environment. Average DCBH seed mass vs. redshift ({\it solid lines}) and their Bondi-Hoyle to Edddington accretion ratio vs. redshift ({\it dashed lines}), for the two zoom-in simulations. The \textsc{zoom16} run starts with a lower average mass as its mass resolution is higher than \textsc{zoom12}. However, the DCBHs in \textsc{zoom16} are able to match the DCBH masses of \textsc{zoom12} quickly along with comparable Bondi to Eddington accretion ratios. This shows that the DCBH growth seen in our simulations is not unusual, but can be reproduced in different numerical setups. Furthermore, the more rapid growth of DCBHs in \textsc{zoom16} during the same range of redshifts as \textsc{zoom12} indicates that \textsc{zoom16} represents an environment that is able to facilitate more efficient DCBH growth. As the \textsc{zoom16} run was initialized from a larger cosmological box, it represents a more biased region of the cosmic web, where the average density is higher to aid SMBH growth.}
    \label{fig:boxcomparison}
\end{figure}

For completeness, we assess the dependence on numerical resolution of our predicted accretion rates in addition to the test done in Fig.~\ref{fig:resolution}, noting that our analysis above has employed the \textsc{zoom12} run. We first consider the gas angular momentum in the halos hosting DCBH seeds to verify that angular momentum is properly conserved. The gas inflow may be artificially enhanced if unphysical numerical dissipation of angular momentum exists. We find that halo gas angular momentum behaves as expected without such numerical dissipation. Fig.~\ref{fig:haloang} shows one example of the expected gas angular momentum evolution near the DCBH seed within the halo of Case C at $z=4.76$. We also compare the gas angular momentum against simple analytical models assuming centrifugally-supported motion, and they generally agree with the simulation results. Furthermore, in Fig.~\ref{fig:boxcomparison}, we now compare the average DCBH seed masses and Bondi-Hoyle to Eddington accretion rate ratios for \textsc{zoom12} and \textsc{zoom16} (see Table~\ref{simtable}).  
At high redshifts, the \textsc{zoom16} run has lower average masses than \textsc{zoom12}. This is because \textsc{zoom16} has a higher mass resolution, so that its gas particle masses are smaller than that of \textsc{zoom12}, and the DCBH seed masses formed from the gas particles are initially lower in turn. However, the Bondi to Eddington accretion ratios are comparable between the two runs, and the DCBH masses in \textsc{zoom16} quickly catch up to \textsc{zoom12}. Thus, the DCBH masses can be reproduced in simulations with different set-ups. We further point out that \textsc{zoom16} was only run up to $z\sim9$ due to computational cost.

\subsection{Conditions affecting AGN growth}\label{conditions}
\begin{figure*}[htb!]
    \centering
    \includegraphics[width=\textwidth]{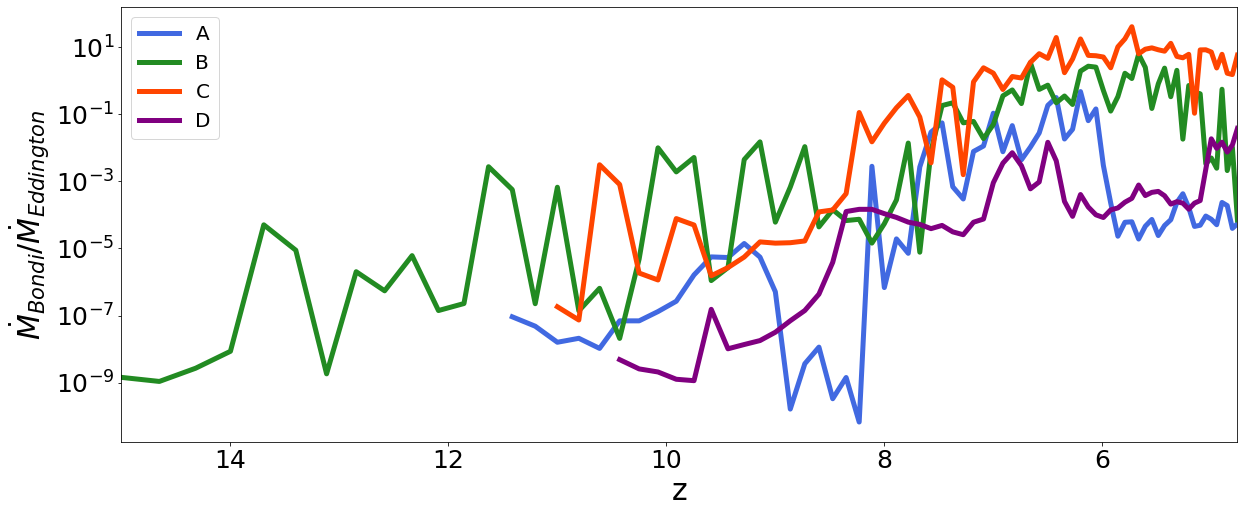}
    \caption{Environment in vicinity of accreting SMBHs. Ratio between the calculated Bondi-Hoyle accretion rate, $\dot{M}_{\rm Bondi}$, to the Eddington value across redshifts for the AGN cases shown in Fig~\ref{fig:massplot}. In general, Bondi-Hoyle accretion rates are lower than the Eddington limit. At later times, when $M_{\rm BH}$ has increased, the Bondi rates may get near or above the Eddington rates. Cases B and C, which reach higher ratios compared to cases A and D, also exhibit continuous growth in Fig.~\ref{fig:massplot}, while B and C show intermittent growth. Therefore, as the Bondi rate depends on the gas properties near the black hole, the presence of dense and cold gas enables efficient growth.}
\label{fig:bondiratio}
\end{figure*}

To understand under what environmental conditions DCBHs can efficiently grow to form the observed high-redshift AGN, we first examine the large-scale cosmological environment the black hole seeds reside in. Again considering Fig.~\ref{fig:boxcomparison}, as stated in Section~\ref{sec:justify}, DCBH masses in \textsc{zoom16} quickly catch up to \textsc{zoom12} despite \textsc{zoom16} starting with a smaller initial DCBH seed mass due to the lower gas particle mass in the higher-resolution run. This indicates that \textsc{zoom16} represents an environment that supported faster DCBH growth than \textsc{zoom12} in the same time period, focused on a more biased region of the Universe with higher average density than \textsc{zoom12}. Efficient SMBH growth at early times thus will be easier to achieve in more biased regions of the cosmic web,  as those regions will likely contain the dense and cold gas environments needed for efficient black hole seed accretion and growth. 

Next, we consider the dark matter host halos for DCBH growth shown in Fig.~\ref{fig:massplot}. The overall halo mass and the halo gas mass does not seem to be strongly correlated with the AGN growth. Case A has the highest halo mass and gas mass out of all the DCBH hosts shown, but experiences a smaller growth compared to the other cases. Thus, while there is ample gas available in the surrounding region where the black hole seed is located in, having more fuel by itself does not facilitate higher black hole growth. The black hole seed will need to efficiently accrete the available gas as well to grow.

We therefore study the environment closer to the black hole seed to see how such efficient accretion may occur. We find that the gas inflow rates near the black hole $(\lesssim100$ pc) do not correlate strongly with the mass accretion rate. This is because the black hole sink particle is not stationary, such that the instantaneous SMBH accretion rate is dependent on local variations of gas temperature and density, as determined by our gas swallow model. In Fig.~\ref{fig:bondiratio}, we plot the ratio between the Bondi-Hoyle accretion rate and the Eddington value, calculated for the conditions at the given redshift for the SMBHs. Note that these accretion rates are calculated in post-processing using the simulated black hole masses and are distinct from the actual accretion rate for the SMBHs in the simulation (Fig.~\ref{fig:dcbhaccretion}), which employs the swallowing prescription (Section~\ref{sec:bhacc}). The Bondi-Hoyle accretion rate depends on the gas properties in the vicinity of the black hole. Thus, if the post-processed Bondi-Hoyle rate is near or larger than the post-processed Eddington accretion rate, it can be inferred that the AGN environment can sustain high accretion rates. Cases B and C, which show nearly continuous growth in Fig.~\ref{fig:massplot}, reach environmental conditions that result in Bondi accretion rates that are higher than Eddington at later times. In contrast, cases A and D, which show periods of stalled growth, do not encounter the conditions for such high Bondi accretion (see Fig.~\ref{fig:dcbhaccretion}). High Bondi-Hoyle accretion rates require dense, cold gas near the black hole (see Equ.~\ref{bondi}), which cases B and C exhibit, in difference from cases A and D. 

\begin{figure*}[htb!]

\gridline{
 \fig{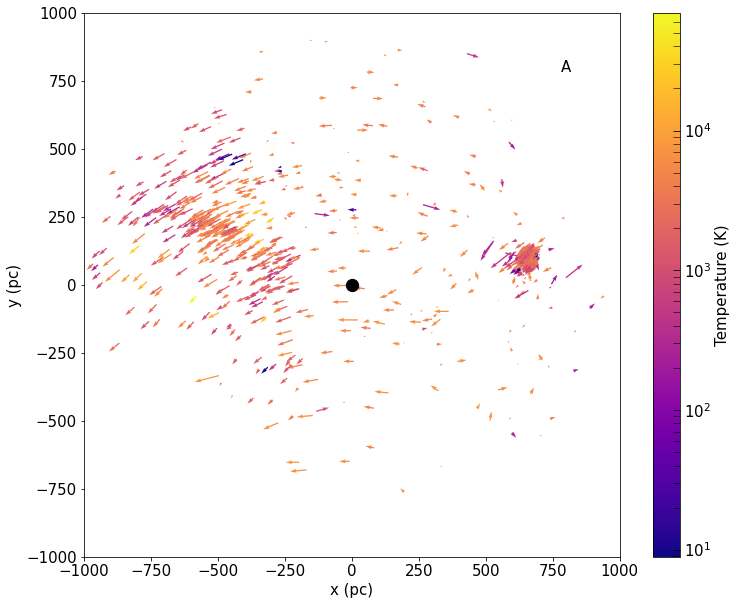}{0.5\textwidth}{}
 \fig{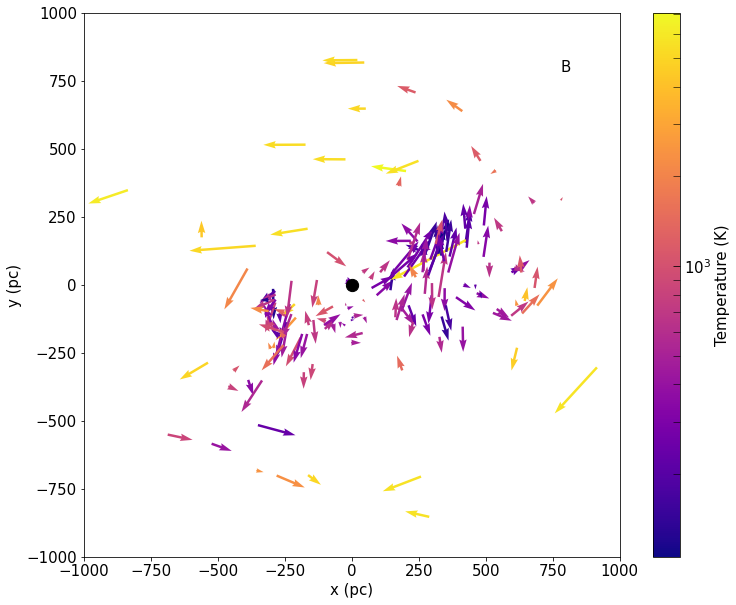}{0.5\textwidth}{}
}
\vspace{-0.50cm}
\gridline{
 \fig{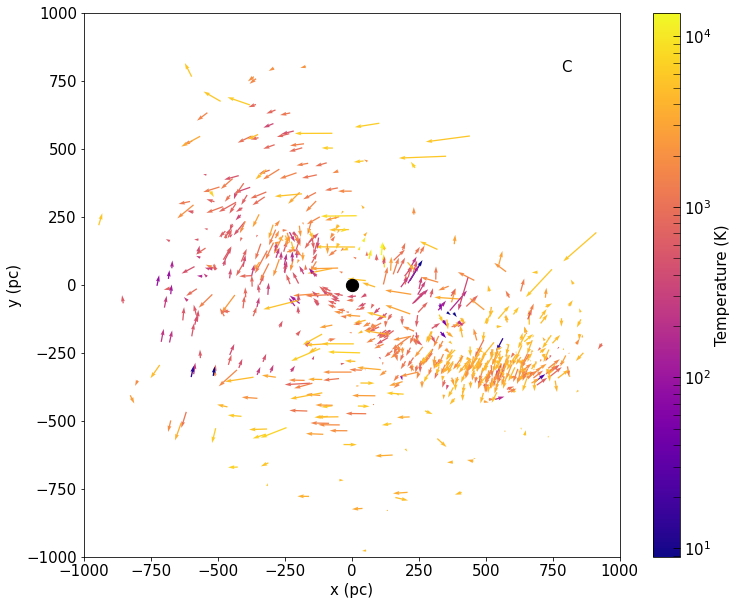}{0.5\textwidth}{}
 \fig{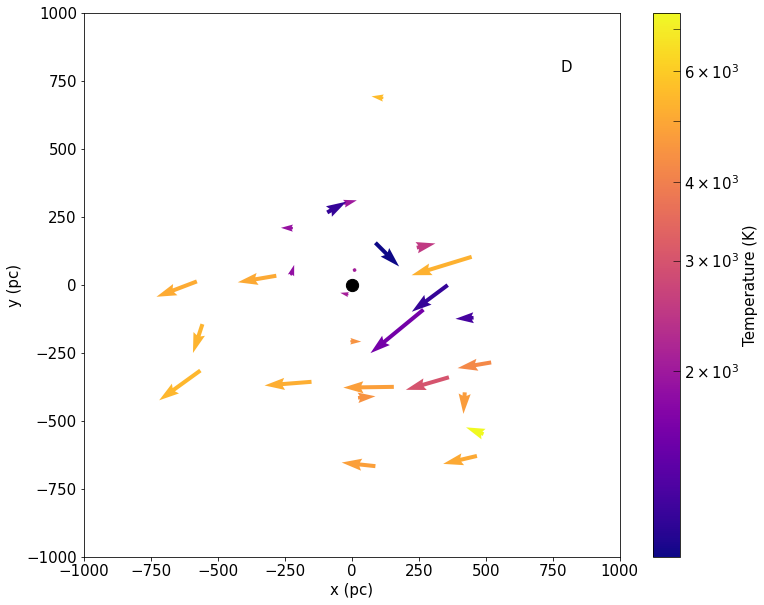}{0.5\textwidth}{}
}
\caption{Velocity field of the gas particles near the DCBH seeds of Fig.~\ref{fig:massplot} at $z\sim6$ in the $xy$ plane. The black dot marks the black hole position and the arrows the moving gas particles. The arrow lengths indicate the velocity magnitudes, scaled to the average velocity and the number of gas particles in each plot. The arrow colors represent the gas temperature in Kelvins. Generally, gas particles both fall into and move away from the black hole seeds with temperatures around $\sim10^3$ K, lower than the stellar feedback heated gas temperature of $\sim10^4$ K. Such conditions are expected as DCBHs form in cold environments. When more infalling gas particles exist, the seed black holes grow as in B, C, while when little to no gas particles exist near the black hole as in D, or when outflowing gas dominates as in A, the black hole growth stalls. Furthermore, the gas particle temperatures of B and C are on average lower than that of A and D. Case B also clearly shows colder gas inflow near the black hole. Thus, black hole growth is enhanced in cold environments with inflowing gas present.
\label{fig:gasvel}}
\end{figure*}

We further explicitly examine the gas particle distribution near the DCBH seeds. Fig.~\ref{fig:gasvel} shows the velocity field of the gas particles and their temperatures near the DCBH seeds in Fig.~\ref{fig:massplot} at $z\sim6$. Generally, gas inflows and outflows both exist near the black hole with temperatures around $10^3$\,K. To grow efficiently, there needs to be an inflow of high density gas. Cases B and C show such conditions, where gas rotates nearby, with a component that is inflowing towards the back hole. However, if there is a strong outflow present as in case A, where the gas is streaming near the black hole, or if the black hole resides in low-density gas, as in case D, with only a few gas particles nearby compared to the other cases, the growth is reduced. 

In addition, the black holes will more efficiently accrete colder gas, in line with the average gas temperatures of cases B and C being lower than that of A and D. Case B further shows cold gas inflow in the vicinity of the black hole which will promote its growth. These results are consistent with Fig.~\ref{fig:dcbhaccretion} and \ref{fig:bondiratio}, where B and C showed more consistent/higher accretion rates compared to A or D. Moreover, B and C were the cases where the black hole mass dominated over stellar mass. Smaller stellar mass thus could facilitate a denser and colder environment, as stellar feedback heats the cold gas and reduces accretion \citep[e.g.,][]{JohnsonBromm2007,Jeon2023}. Therefore, efficient accretion for the AGN population occurs in more biased and dense regions of the Universe. The DCBH seeds in our simulations generally do reside in such environments. This is consistent with the formation criteria for a DCBH seed, including the requirement of cold, dense gas conditions. Thus, the formation mechanism for the heavy DCBH seeds can also promote their efficient growth, at least initially \citep[e.g.,][]{Becerra2018a}.

\subsection{Reproducing \textit{JWST} Observations}
\begin{figure}[htb!]
    \centering
    \includegraphics[width=0.5\textwidth]{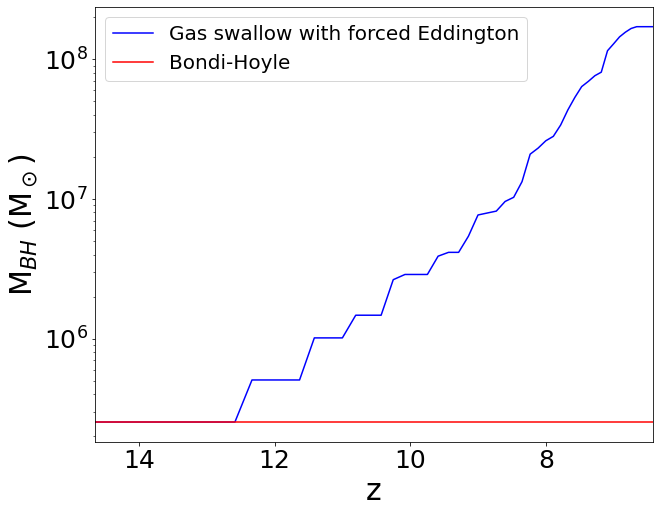}
    \caption{Comparison of manually inserted DCBH seed growth, with an initial mass of $10^5$\,M$_{\odot}$, in the same halo but with different accretion models. Under the fiducial Bondi-Hoyle subgrid model, the DCBH seed does not grow significantly \citep{Jeon2023} and cannot reach $10^{7-8}$ M$_\odot$ by $z\sim8$ to match observations \citep[e.g.,][]{Larson2023}. In contrast, the gas swallow accretion model employed in this paper, designed to follow the Eddington accretion rate when possible, is able to grow the DCBH seed to reproduce observed masses. Therefore, massive seeds and efficient accretion both are needed to reproduce the new population of high-redshift AGN, as observed by \textit{JWST}.}
    \label{fig:insertcompare}
\end{figure}

With the gas swallow model, operating in the high-density and cold gas environments the DCBH seeds reside in, we are able to reproduce the AGN newly observed by the \textit{JWST}, as well as some lower-mass quasars observed previously with masses $\sim10^7-10^8$ M$_\odot$ at $z\sim7-8$, as can be seen in Fig.~\ref{fig:massplot}. To do so, we require that the SMBHs grow at rates larger than the standard Bondi-Hoyle accretion rate. Fig.~\ref{fig:insertcompare} shows two runs where we explicitly insert a DCBH seed of $10^{5}$ M$_\odot$ in the same halo at $z=15$, using the Bondi-Hoyle accretion model for one, and the gas swallowing model used in this paper for the other. As is evident, Bondi-Hoyle accretion is inefficient in growing the black hole \citep{Jeon2023}. In contrast, our gas swallow model does grow the seed exponentially by enforcing Eddington accretion when possible. Observations have detected AGN at $z\sim8$ with SMBH masses of $10^{7-8}$ M$_\odot$, which the gas swallow model can reproduce but the Bondi-Hoyle model cannot. Thus, our simulations suggest that a heavy seed must accrete efficiently to produce massive AGN so early in cosmic history, implicating rarer and more extreme conditions that produce the required high accretion rates (see Section~\ref{conditions}).

Furthermore, Fig.~\ref{fig:bondiratio} shows that the post-processed Bondi-Hoyle accretion proceeds at a rate that is generally much lower than the Eddington value. The Bondi rate starts to match or even exceed the Eddington rate at later times, due to the higher SMBH masses reached at those times. However, if the black holes grew under Bondi-Hoyle subgrid model instead of the gas swallow model, rates would have stayed sub-Eddington with minimal mass growth, as shown in Fig.~\ref{fig:insertcompare}. Thus, to form the high-redshift AGN observed by \textit{JWST}, we require heavy DCBH seeds that grow efficiently following Eddington accretion. 


\subsection{Overmassive SMBHs}\label{sec:overmassive}

We now examine the evolution of the DCBH seeds and their environment together.  All cases in Fig.~\ref{fig:massplot}, including when the stellar mass is always dominant over the DCBH, represent overmassive black holes where the black hole to stellar mass ratio is larger than what is observed locally, similar to a subset of recent \textit{JWST} observations of high-redshift AGN \citep[e.g.,][]{Larson2023,Bogdan2023,Kokorev2023,Pacucci2023}. 

\begin{figure*}[htb!]
\gridline{
 \fig{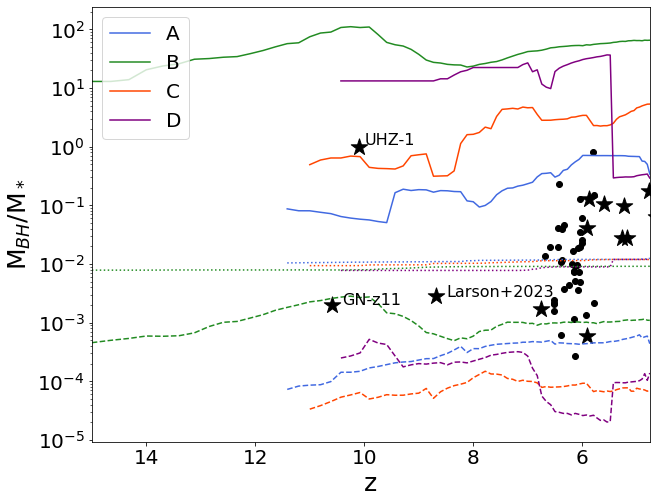}{0.5\textwidth}{}
 \fig{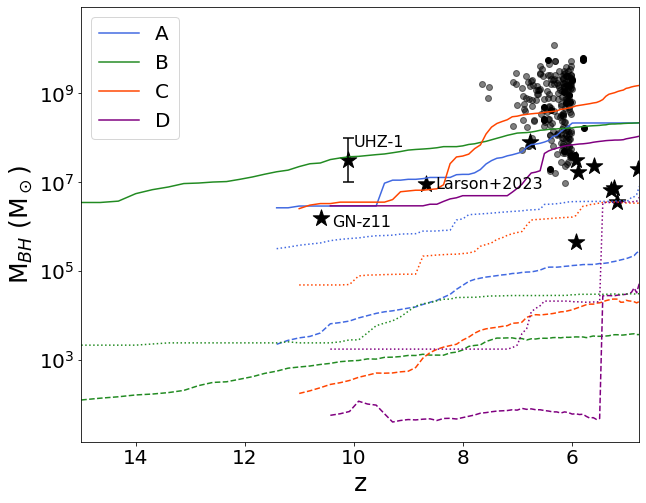}{0.5\textwidth}{}
 }  \caption{Properties of the first SMBHs. {\it Left panel:} Black hole mass to halo stellar mass ratio vs. redshift. {\it Right panel:} Masses of the growing DCBH seeds, shown in Fig.~\ref{fig:massplot}, across redshifts. We also plot in dashed lines the expected black hole mass given the halo mass from the local empirical $M-\sigma$ relation \citep{Evrard2008,Kormendy2013}. The dotted lines represent the SMBH mass to host stellar mass ($M_{\rm BH}-M_{\ast}$) relation derived from \textit{JWST}-observed galaxies at $z\sim4-7$ \citep{Pacucci2023}. The black stars represent select high-redshift AGN observed with \textit{JWST} \citep{Larson2023,Bogdan2023,Maiolino2023_2}, and the black dots show a selection of observed high-redshift quasars \citep{Wang2010,Willott2017,Decarli2018,Izumi2018,Pensabene2020,Inayoshi2020}. We note that the $z\sim10$ AGN in UHZ1 \citep{Bogdan2023} has a similar ratio as our simulated halos, whereas the \citet{Larson2023} source at $z\sim8$ does not. In contrast, both black hole masses are reproduced by our simulations. Generally, our simulated halos are overmassive until $z\sim5$, but no such systems have been observed at these redshifts. This may reflect the limitations of our simulation box sizes, missing the larger-scale modes that contribute to structure formation at lower redshifts. However, at $z\sim10$, when the large-scale effects are not yet prominent, select observations match our simulations. Since the simulated ratios for the DCBH seeds are all above the local relation extrapolated to these higher redshifts, processes at later cosmic times, such as vigorous star formation and/or mergers, will be necessary for these systems to evolve to the local galaxy populations.}
    \label{fig:overmassive}
\end{figure*}

In Fig.~\ref{fig:overmassive}, we compare the black hole to stellar mass ratio and the masses for the DCBH seeds against the local $M-\sigma$ relation \citep{Evrard2008,Kormendy2013}, as well as against the SMBH to host stellar mass ($M_{\rm BH}-M_{\ast}$) relation derived from galaxies observed by \textit{JWST} at $z\sim4-7$ \citep{Pacucci2023}. The DCBH seeds all lie above the local and the higher-redshift relation. We also plot the black hole to stellar mass ratios and black hole masses for select high-redshift observations of AGN with \textit{JWST} \citep{Larson2023,Bogdan2023}, and of previously observed quasars \citep{Wang2010,Willott2017,Decarli2018,Izumi2018,Pensabene2020,Inayoshi2020}. The intriguing UHZ1 source at $z\sim10$ \citep{Bogdan2023} agrees with the simulated stellar to black hole mass ratios, whereas the \citet{Larson2023} AGN at $z\sim8.7$ approaches the local $M-\sigma$ ratio. However, both AGN masses are reproduced by the simulations. 

Furthermore, the simulated systems remain overmassive until $z\sim5$, whereas no such galaxies have been observed at those redshifts. This may be an issue arising from the limited size of our simulation box. At later times, larger-scale density fluctuation modes from the primordial Universe contribute to structure formation and galaxy evolution, but they cannot be imprinted in the initial conditions of computational boxes of insufficient size. Consequently, at lower redshifts galaxies and their stellar components cannot grow as massive as they would have in the presence of those missing modes. Thus, our halo and stellar masses at later times may be underestimated. However, at higher redshifts ($z\sim10$) when the large-scale contributions are not yet prominent, our simulation box should be adequate to represent the initial growth stages of DCBH seeds. This indicates that the $M-\sigma$ relation is still rapidly evolving at the early time simulated here. Future episodes of star formation and/or mergers with other galaxies (as in the merger at $z\sim5.5$ of Fig.~\ref{fig:massplot}D) may be needed for these objects to evolve to the local relation. Finally, to arrive at the local $M-\sigma$ relation, a mutual feedback-cycle between the components has to be established over sufficiently long periods of time \citep[e.g.,][]{Silk2024}, a process which is not properly captured by our simulations here.

\begin{figure*}[htb!]

\gridline{
 \fig{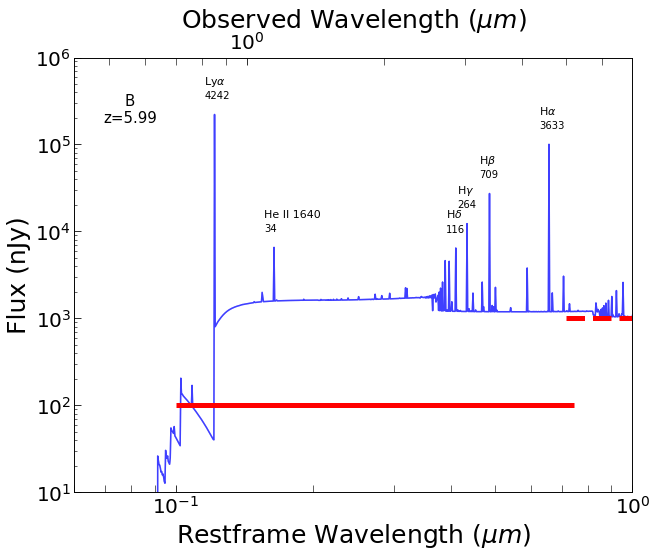}{0.4\textwidth}{}
 \fig{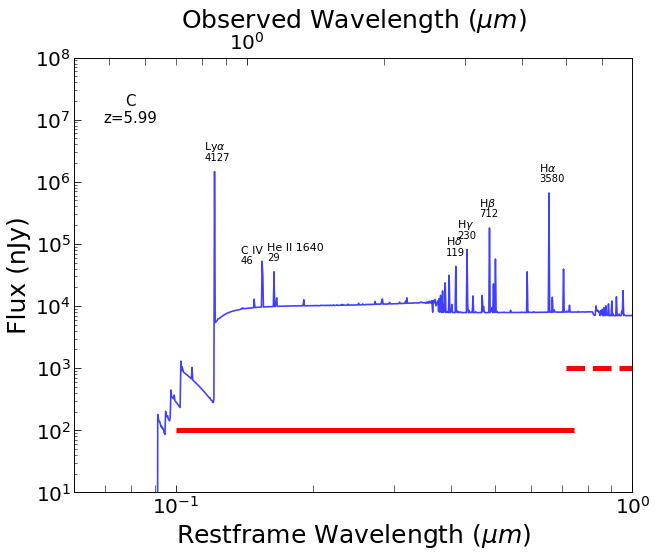}{0.4\textwidth}{}
}
\vspace{-0.50cm}
\gridline{
 \fig{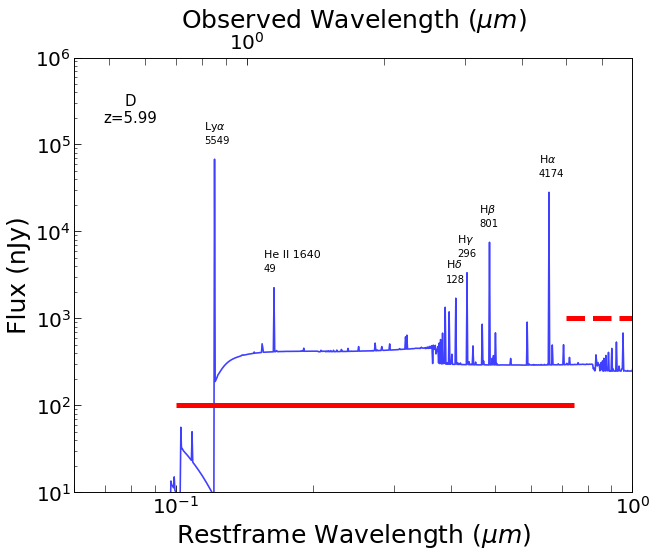}{0.4\textwidth}{}
 \fig{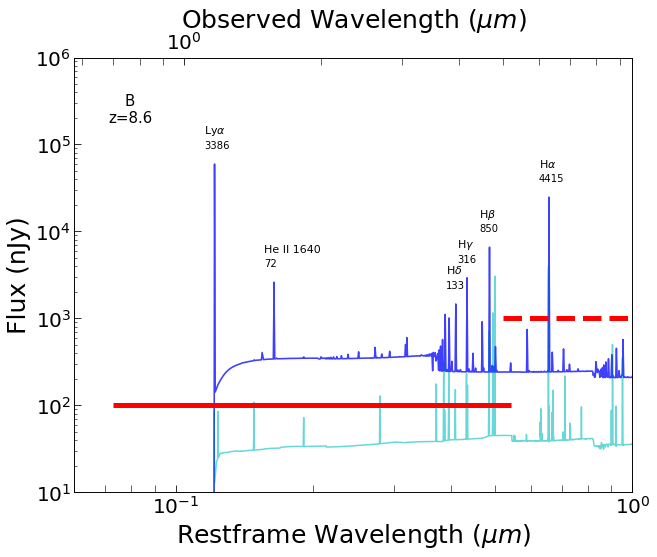}{0.4\textwidth}{}
}
\vspace{-0.50cm}
\gridline{
 \fig{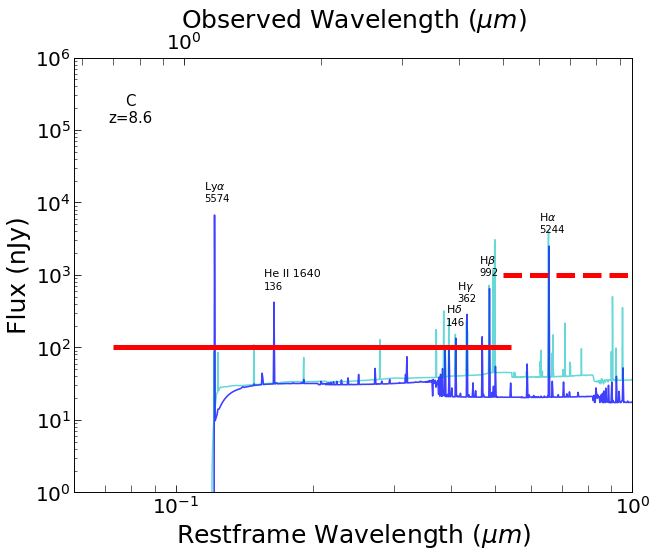}{0.4\textwidth}{}
 \fig{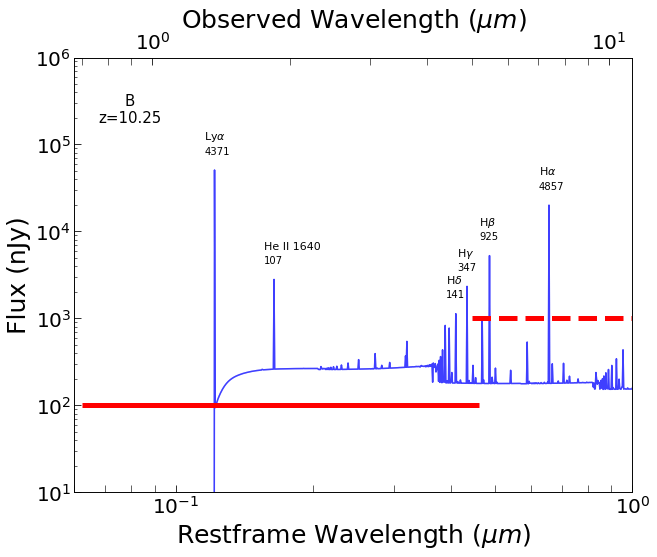}{0.4\textwidth}{}
}
\caption{Predicted spectra of the AGN shown in Fig.~\ref{fig:massplot}, using \textsc{Cloudy} version C23.01 \citep{Chatzikos2023}. For the input continuum, we use the multi-color disk model of \citet{Jeon2014}. We further apply intergalactic medium absorption from \citet{Madau1995,Madau1996}. We show spectra for $z=6$, $z=8.6$, and $z=10.25$, whereas cases without spectra had too small accretion rates to have a meaningful flux at the chosen redshifts. We specifically compare the $z=8.6$ spectrum against the \citet{Larson2023} AGN, whose modeled/inferred continuum spectrum is shown in light blue. We also indicate with the red solid (dashed) line the approximate lower flux limit probed by \textit{JWST} NIRSpec (MIRI) for a signal to noise ratio of 10 with $10^4$\,seconds of integration. The most prominent emission features are the hydrogen and helium lines. This is due to the generally low metallicity of the AGN environment, characteristic of the regions that can facilitate DCBH formation and growth. For the most prominent lines, we add above them their identification and equivalent widths in {\AA}ngstroms. The accreting SMBHs are mostly expected to be observable, consistent with the existing \textit{JWST} observations of select high-redshift AGN.
\label{fig:cloudyobserve}}
\end{figure*}

\section{Observability}\label{sec:observe}

To assess the observability of the simulated AGN, we employ the \textsc{Cloudy} code, version 23.01 \citep{Chatzikos2023}. We further use the multi-color disk spectrum of \citet{Jeon2014}, produced by the accreting SMBHs, as input spectra for \textsc{Cloudy}, normalized to the bolometric luminosity of $L_{\rm BH} = \epsilon_0\dot{M}c^2$. In \textsc{Cloudy}, we define a gas cloud with density, metallicity, and distance from the black hole set to the average values for the gas particles within the black hole kernel. For simplicity, we assume that the gas pressure is constant inside the cloud. \textsc{Cloudy} passes the normalized AGN spectra through the gas cloud, thus producing the reprocessed spectrum. We also apply intergalactic medium absorption to the output spectra, following \citet{Madau1995,Madau1996}. The reprocessed AGN spectrum is then converted to the observable flux that will reach us. Specifically, we obtain spectra for $z=6, 8.6, 10.25$, with the latter two corresponding to the redshifts of the \textit{JWST}-observed AGN in \citet{Larson2023,Bogdan2023}, respectively. 

Fig.~\ref{fig:cloudyobserve} summarizes the resulting spectra. Some AGN cases in Fig.~\ref{fig:massplot} had little to no accretion at the selected redshifts and are thus not shown here. We also indicate the wavelength ranges covered by \textit{JWST} NIRSpec and MIRI (solid and dashed red lines, respectively), as well as their corresponding sensitivity flux limits, assuming a signal to noise ratio of 10 with $10^4$\,seconds of integration. For the spectra at $z=8.6$, we overplot (in light blue) the modeled continuum spectra for one of the \textit{JWST} AGN observations at high-$z$ \citep{Larson2023}. Compared to the simulated AGN spectra, the \citet{Larson2023} spectrum has a weaker Balmer break near 0.36 $\micron$ (rest-frame). This may be due to the source's continuum spectrum being dominated by stars to produce more ionizing photons compared to our simulated spectrum that only includes the radiation from the AGN. For case B, the predicted spectral continuum is much higher than that inferred for \citet{Larson2023}, while for case C, they are comparable. Thus, high-redshift AGN can produce emission comparable to or larger than the stellar component of high-redshift galaxies under favorable accretion conditions, allowing them to be detected with \textit{JWST}. 

Furthermore, the strongest emission lines of the simulated spectra are the hydrogen Lyman and Balmer series and the helium He\,II lines. Other lines are relatively weaker and only a few metal lines are present, such as the C\,IV line between the Lyman-$\alpha$ and He\,II lines at $z\sim6$ for case C. Case C also exhibits an interstellar medium environment with the highest metallicity of $\sim 0.1 $\,Z$_{\odot}$. Such dominance of non-metal emission lines reflects the initially near-pristine chemical environment able to host and efficiently grow DCBH seeds, including the absence of vigorous star formation during the early stages of formation. 

The simulated AGN that are accreting are mostly predicted to be observable even up to $z\sim10$ as fluxes are above the detection threshold at the probed wavelengths, consistent with the \textit{JWST} observations of such systems. Only the case~C continuum at $z=8.6$ lies below \textit{JWST} limits, but strong nebular emission lines, such as the Lyman-$\alpha$ line should still be observable \citep[e.g.,][]{Jeon2019}. In reality, the stellar component of the galaxy will also contribute to the observed spectrum which will make identifying the AGN at high redshift more difficult. However, as stated above, our AGN spectra exhibit flux levels comparable to the stellar spectrum of the observed \citet{Larson2023} system, and all our systems are overmassive, with the AGN having comparable masses as the stars across redshifts (see Fig.~\ref{fig:overmassive}). They will most likely behave as the overmassive systems that have already been observed \citep[e.g.][]{Bogdan2023,Kokorev2023,Pacucci2023}.



\section{Summary and Conclusions}\label{sec:conclude}
Using cosmological simulations and the enhanced black hole growth model through the gas swallow methodology, we have reproduced the high-redshift ($z\sim8-9$) massive AGN (M$_{\rm BH}\sim10^7-10^8$ M$_\odot$) that have been observed with the \textit{JWST}. The gas swallow model imposes growth of the massive seeds at close to the Eddington limit, modulated by the availability of gas within the host halo. Such efficient accretion mode reflects physics on smaller scales, such as magnetic fields and nonuniform density \citep{Jiang2019_2,Davis2020}, that are unresolved in our simulations. We further investigate what environmental conditions allow for such efficient gas accretion and black hole growth. More biased and dense regions of the cosmic web promote the overall growth of the AGN population, whereas individual AGN growth is enhanced by cold and dense local environments with gas inflows towards the black hole. Such local environments may naturally arise in regions that enable DCBH seed formation. 

We compare our simulated AGN growth trajectories against various observations. We are able to reproduce the black hole masses of high-redshift AGN observed with \textit{JWST}, and also the lower-mass regime of high-redshift quasars observed prior to \textit{JWST}. In addition, we produce the observed overmassive systems \citep[e.g.,][]{Bogdan2023,Kokorev2023,Pacucci2023}, where the stellar mass of the halo is comparable to the SMBH mass. Such overmassive configurations are specifically predicted for systems that can form heavy DCBH seeds \citep{Wise2019,Agarwal2013}. We derive spectra for these systems using the \textsc{CLOUDY} code \citep{Chatzikos2023}, demonstrating that they are in most cases observable with the \textit{JWST}. This is in difference to black hole growth trajectories within the first galaxies that start with `light', stellar-remnant seeds and accrete at sub-Eddington rates, where the resulting low-luminosity AGN typically remain below current detection thresholds \citep[e.g.,][]{Jeon2023}.

Our study is subject to a number of caveats and simplifications, requiring future work to address them. As discussed above, the limited size of our simulation boxes reduces accuracy in reproducing halo and structure growth at lower redshifts, given the absence of large-scale perturbation modes that cannot be represented in our initial conditions. Consequently, the overmassive systems encountered in our simulations generally do not evolve closer to the canonical $M-\sigma$ relation, but instead remain overmassive even by $z\sim5$, whereas no such objects have been observed at these redshifts. 

We further note that the number density of DCBHs in our simulation might be not fully realistic. We find around 10 DCBH seeds in the zoom-in regions for parent simulations of linear sizes 12$h^{-1}$ and 16$h^{-1}$ comoving Mpc. Previous works have found much smaller predicted DCBH number densities \citep{Habouzit2016,Mayer2019,Sassano2021}. We employ a simple model for DCBH seeding, inserting a fully-formed DCBH when a group of gas particles satisfies the density and chemistry criteria for DCBH formation, and converting the densest gas to a DCBH sink particle. Therefore, it might not fully capture the DCBH formation process \citep[e.g.,][]{Hosokawa2013,Barrow2018}, forming too many DCBHs for a given box size. However, in this study we were concerned with the growth and evolution of DCBHs, and did not focus on the overall DCBH population. The enhanced formation instead resulted in an ensemble of independently evolving DCBH cases in a computationally feasible cosmological simulation. The `over-seeding' is thus not a crucial problem for this work, but will have to be addressed in future studies that aim to derive the global impact these massive, accreting seeds might have on the evolution of the early Universe. 

How was the Universe able to form SMBHs so early in its history \citep{Schneider2023}? This conundrum is part of the larger challenge to understand the stunning prevalence of massive structures and galaxies in the first few 100 million years after the Big Bang, as recently revealed by \textit{JWST} \citep{MBK2023}. Furthermore, the growth of galaxies and their central SMBHs is intricately linked through a mutual feedback cycle \citep[e.g.,][]{SilkRees1998}, which may well have operated differently in the high-redshift Universe \citep[e.g.,][]{Dekel2023}. Answering these fundamental questions, and more generally characterizing the evolution and impact of the first SMBHs, will rely on future, ultra-deep observations with the \textit{JWST}, as well as with complementary frontier facilities, including next-generation gravitational wave observatories \citep{Amaro2023}, the \textit{Roman Space Telescope} \citep{Mosby2020}, and X-ray observatories like \textit{Athena} \citep{Barret2013} and \textit{Lynx} \citep{Gaskin2019}. Astronomy is about to open up a window into the very onset of the cosmic co-evolution of galaxies and black holes, and we are already discovering tantalizing hints into the infancy of the most massive objects in the Universe.

\appendix

\section{Impact of Black Hole Feedback on Mass Growth}
\begin{figure}[!htb]
    \centering
    \includegraphics[width=0.8\textwidth]{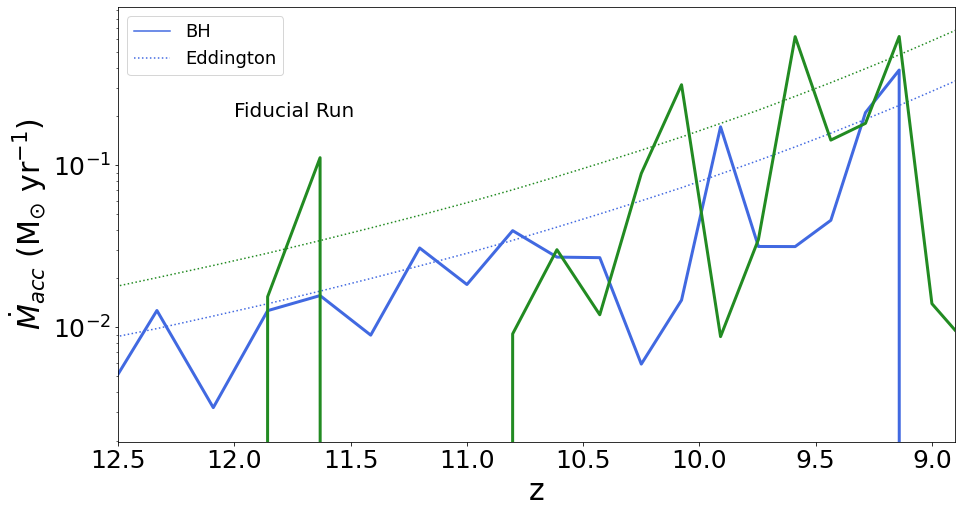}
    \includegraphics[width=0.8\textwidth]{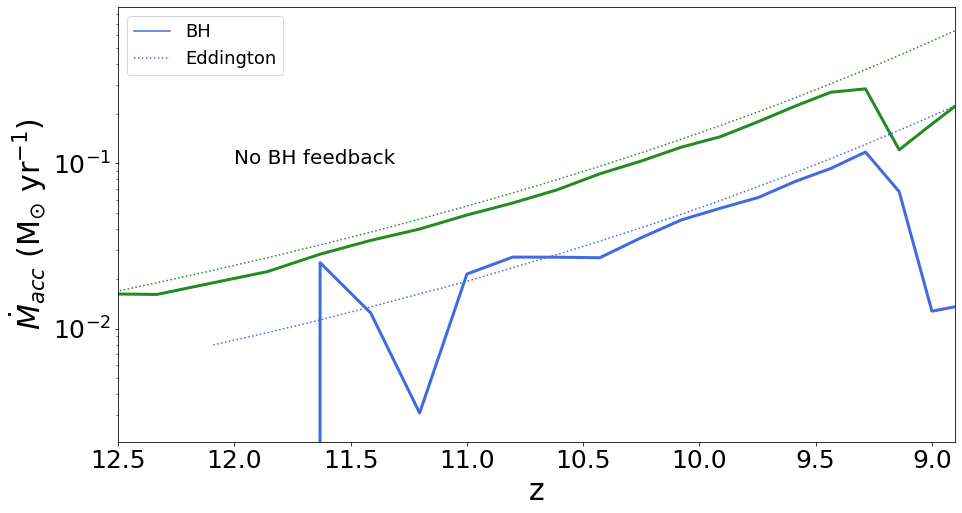}
    \caption{Mass accretion rates for select DCBH seeds (\textit{solid lines}) and their respective Eddington rates (\textit{dotted lines}) for the fiducial run with black hole thermal feedback and the run without such feedback. Here, colors ({\it blue and green}) represent different individual DCBH seeds chosen at random from the simulations. In general, DCBHs in both runs follow Eddington accretion as designed by our algorithm, but without BH feedback, the accretion rate closely tracks the Eddington limit for extended periods, while BH growth shows stronger fluctuations in the fiducial run with BH feedback. Therefore, black hole feedback can partially suppress BH accretion, acting to regulate BH growth.}
    \label{fig:macc_edd_main}
\end{figure}

To assess the effect of black hole feedback on its accretion and growth, we have performed a simulation run without including black hole thermal feedback (see Section~\ref{sec:bhacc}), comparing with the fiducial run where this feedback is present. Fig.~\ref{fig:macc_edd_main} shows the accretion rates for two randomly selected DCBH seeds formed in each of the simulation boxes, compared to their respective Eddington rates. The accretion rate generally tracks the Eddington rate as imposed by our swallowing prescription (see Section \ref{sec:bhacc}), and without BH feedback, the two rates closely coincide for extended periods. Thus, black hole feedback reduces accretion onto the black hole by heating the nearby gas, and works to regulate bursty black hole growth following feedback cycles, as high accretion will lead to stronger feedback and a period of decreased growth, while low accretion will result in weaker feedback boosting subsequent mass growth. 

%

\begin{acknowledgments}
We thank Chris Richardson for his assistance in implementing and running \textsc{Cloudy}. The authors acknowledge the Texas Advanced Computing Center (TACC) for providing HPC resources under allocation AST23026. BL gratefully acknowledges the support of the Royal Society University Research Fellowship and the funding from the Deutsche Forschungsgemeinschaft (DFG, German Research Foundation) under Germany's Excellence Strategy EXC 2181/1 - 390900948 (the Heidelberg STRUCTURES Excellence Cluster). 
\end{acknowledgments}







\bibliography{ms}{}
\bibliographystyle{aasjournal}



\end{document}